\documentclass[a4paper,oneside,final,notitlepage,onecolumn,12pt]{article}

\usepackage[utf8]{inputenc}
\usepackage[british]{babel}

\usepackage{amsmath}
\usepackage{amssymb}
\usepackage{authblk}
\usepackage{mathrsfs}

\usepackage{outlines}

\usepackage{caption}
\usepackage{subcaption}
\usepackage{graphicx}
\usepackage{overpic}
\usepackage{amsmath,accents}
\allowdisplaybreaks 
\usepackage{float}
\usepackage[draft]{showlabels}

\newcommand{\interior}[1]{\accentset{\smash{\raisebox{-0.12ex}{$\scriptstyle\circ$}}}{#1}\rule{0pt}{2.3ex}}
\newcommand{\instar}[1]{\accentset{\smash{\raisebox{-0.12ex}{$\scriptstyle\star$}}}{#1}\rule{0pt}{2.3ex}}
\newcommand{\indot}[1]{\accentset{\smash{\raisebox{-0.1ex}{$\scriptstyle{\bigcdot}$}}}{#1}\rule{0pt}{2.3ex}}
\selectfont

\usepackage{hyperref}
\hypersetup
{
    colorlinks=true,
    linktocpage=true,
    linkcolor={red},
    anchorcolor={black},
    citecolor={green},
    menucolor={red},
    unicode=true,
    pdftitle={Is it possible to construct asymptotically flat initial data using the evolutionary forms of the constraints?},
    pdfauthor={Károly Z. Csukás, István Rácz}
}

\newcommand{\aaa}{\mathbf{a}}
\newcommand{\bb}{\mathbf{b}}
\newcommand{\bbb}{\overline{\mathbf{b}}}
\newcommand{\NN}{\mathbf{N}}
\newcommand{\Kcqq}{\Kc_{qq}}
\newcommand{\Kcqqb}{\Kc_{q\overline{q}}}
\newcommand{\KK}{\mathbf{K}}
\newcommand{\bgKK}{{}^{\scriptscriptstyle(\!0\!)}\hspace{-1.2pt}\KK}
\newcommand{\dKK}{{}^{\scriptscriptstyle(\!\Delta\!)}\hspace{-1.2pt}\KK}
\newcommand{\kk}{\mathbf{k}}
\newcommand{\kkb}{\overline{\mathbf{k}}}
\newcommand{\dkk}{{}^{\scriptscriptstyle(\!\Delta\!)}\hspace{-1.2pt}\kk}

\newcommand{\bgkk}{{}^{\scriptscriptstyle(\!0\!)}\hspace{-1.2pt}\kk}

\newcommand{\Nh}{\widehat{N}}
\newcommand{\bgNh}{{}^{\scriptscriptstyle(\!0\!)}\hspace{-3pt}\Nh}
\newcommand{\dNh}{{}^{\scriptscriptstyle(\!\Delta\!)}\hspace{-3pt}\Nh}
\newcommand{\NNt}{\mathbf{\NN}}
\newcommand{\kkappa}{\boldsymbol{\kappa}}
\newcommand{\bgkkappa}{{}^{\scriptscriptstyle(\!0\!)}\hspace{-1.2pt}\kkappa}

\newcommand{\Kstar}{\instar{K}}
\newcommand{\Nt}{\mathbf{N}}
\newcommand{\Ntb}{\overline{\mathbf{N}}}
\newcommand{\ethb}{\overline{\eth}}
\newcommand{\dd}{\mathbf{d}}
\newcommand{\AAA}{\mathbf{A}}
\newcommand{\BB}{\mathbf{B}}
\newcommand{\CC}{\mathbf{C}}
\newcommand{\AAAb}{\overline{\mathbf{A}}}
\newcommand{\BBb}{\overline{\mathbf{B}}}
\newcommand{\CCb}{\overline{\mathbf{C}}}
\newcommand{\ff}{\mathbf{f}}
\newcommand{\FF}{\mathbf{F}}
\newcommand{\Rh}{\widehat{R}}
\newcommand{\Kc}{\interior{\KK}}
\newcommand{\Kh}{\widehat{K}}
\newcommand{\Rthree}{{}^{\scriptscriptstyle(\!3\!)}\!R}

\renewcommand{\imath}{\mathrm{i}}
\renewcommand{\Re}{\mathfrak{Re}}
\renewcommand{\Im}{\mathfrak{Im}}

\usepackage[normalem]{ulem}
\usepackage{color}

\newcounter{mnotecount}

\newcommand{\mnotex}[1]
{\protect{\stepcounter{mnotecount}}$^{\mbox{\footnotesize $\bullet$\themnotecount}}$
	\marginpar{\color{red}
		\raggedright\tiny\em
		$\!\!\!\!\!\!\,\bullet$\themnotecount: #1} }

\makeatletter
\newcommand*{\bigcdot}{}
\DeclareRobustCommand*{\bigcdot}{%
	\mathbin{\mathpalette\bigcdot@{}}%
}
\newcommand*{\bigcdot@scalefactor}{.5}
\newcommand*{\bigcdot@widthfactor}{1.15}
\newcommand*{\bigcdot@}[2]{%
	\sbox0{$#1\vcenter{}$}
	\sbox2{$#1\cdot\m@th$}%
	\hbox to \bigcdot@widthfactor\wd2{%
		\hfil
		\raise\ht0\hbox{%
			\scalebox{\bigcdot@scalefactor}{%
				\lower\ht0\hbox{$#1\bullet\m@th$}%
			}%
		}%
		\hfil
	}%
}
\makeatother

\usepackage[backend=biber,style=nature,sorting=none,backref=true]{biblatex}
\title{Is it possible to construct asymptotically flat initial data using the evolutionary forms of the constraints?}
\author[,1,2]{Károly Csukás \footnote{E-mail address:{\tt csukas.karoly@wigner.hu}}}
\author[,1]{István Rácz \footnote{E-mail address:{\tt racz.istvan@wigner.hu}}}

\affil[1]{\small Wigner RCP, H-1121 Budapest, Konkoly Thege Mikl\'{o}s \'{u}t  29-33, Hungary}
\affil[2]{\small The University of Mississippi, University, Mississippi 38677, USA}

\begin{document}
  \maketitle

  \begin{abstract}
  	Near-Kerr black hole initial datasets are constructed by applying either the parabolic-hyperbolic or the algebraic-hyperbolic form of the constraints. In both cases, strongly and weakly asymptotically flat initial datasets with desirable falloff rates are produced by controlling only the monopole part of one of the freely specifiable variables. The viability of the applied method is verified by numerically integrating the evolutionary forms of the constraint equations in the case of various near-Kerr configurations.
  \end{abstract}

	\bigskip

  \tableofcontents

  \section{Introduction}

	For more than seven decades, the conformal or elliptic method, introduced by Lichnerowicz and York \cite{Lichnerowicz:1944, Lichnerowicz:1952, York:1972prl, York:1975aihp}, was almost the only\,\footnote{There have been attempts to find other means of solving the constraints (see e.g. \cite{Stachel:1978, Bishop:2004}), but none of them has proved to be a viable alternative to the conformal method.} means to solve the constraints of general relativity. Recently, in a series of papers \cite{Racz:2014gea, Racz:2014jra, Racz:2015mfa}, two alternative evolutionary formulations of the Einstein constraint equations were also introduced. It is well known that by specifying a suitable boundary value problem within the elliptic method, it is straightforward to arrange a setup capable of investigating the existence of asymptotically flat solutions to the constraint equations. By contrast, it is not evident if there is a way to guarantee the existence of asymptotically flat solutions to the constraint equations when either of the novel parabolic-hyperbolic or algebraic-hyperbolic forms of the constraints is used. The main issue is whether controlling merely the freely specifiable part of the data, along with the ``initial data'' relevant for the constraint variables---which are given only on one of the level sets foliating the conventional initial data surface---is sufficient to guarantee appropriate falloff behavior of all the geometric fields involved in the constraints.

	By their inherent nature, the constraints involve more variables than equations; thereby, they always form an underdetermined system. For instance, there are only four constrained variables, regardless of the method applied, concerning the geometrical content of a three-dimensional initial dataset in general relativity. These are the ones restricted by the Hamiltonian and momentum constraints \cite{Choquet-Bruhat:2014okh, Wald:1984rg}. In contrast, the remaining eight variables are freely specifiable. Whenever one is looking for asymptotically flat solutions located in a small neighborhood of a known one, it is very tempting to choose all the freely specifiable variables to coincide with the corresponding variables of the known asymptotically flat solution. This strategy was applied by Beyer {\it et al.} \cite{Beyer:2017njj, Beyer:2019kty} in their pioneering studies of the asymptotic behavior of solutions to the evolutionary form of vacuum constraints. More concretely, they used data on a time slice of the Kerr-Schild form of a single or superposed binary Schwarzschild black holes, and they altered only the ``initial data''---these were also read off on one of the foliating level surfaces---for some of the constrained variables  \cite{Beyer:2017njj, Beyer:2019kty}. Their investigations demonstrated that for this choice of the freely specifiable data---we referred to these in \cite{Csukas:2019qco} as ``strictly near Schwarzschild'' configurations---the solutions to the evolutionary form of the constraints, apart from the seed solution, cannot be asymptotically flat. More concretely, in the strictly near-Schwarzschild setup, neither the algebraic-hyperbolic formulation in the single Schwarzschild case \cite{Beyer:2017njj} nor the parabolic-hyperbolic formulation in the single and binary Schwarzschild black hole case \cite{Beyer:2019kty} allow suitable falloff for one of the constrained variables, $\KK$, that is nothing but the trace of the tensorial projection of the extrinsic curvature [for its definition see Eq. \eqref{eq:traceKK} below].\footnote{Strictly speaking, $\KK$ is not the only variable that violates the desired decay rate, but given the spherically symmetric constraint equations, it can be argued that the slow decay rate of the other problematic variables can always be traced back to the behavior of $\KK$.} These findings were confirmed in \cite{Csukas:2019qco}. Note also that our mode-by-mode (numerical) analysis also allowed us to conclude that only the monopole part of $\KK$ violates the desired decay rate \cite{Csukas:2019qco}. Accordingly, this monopole part is the only mode that gets in the way of obtaining asymptotically flat solutions to the evolutionary form of the constraints in the strictly near-Schwarzschild case \cite{Csukas:2019qco}.

	It is important to emphasize that in the above-discussed investigations, the enormous freedom we have in choosing the freely specifiable part of the data was ignored entirely. Based on this observation and to ensure a suitable falloff rate for the constrained variable, $\KK$, Beyer {\it et al.}\,\cite{Beyer:2020zlo} invented a method that applies to the parabolic-hyperbolic form of the constraints, by setting  $\kkappa$ ---which is one of the freely specifiable variables, and which is the scalar part of the extrinsic curvature [see Eq. \eqref{eq:decompKK} below]---to be proportional to $\KK$. Note that by assuming $\kkappa = \mathcal{R}\,\KK$, the principal part of the constraints' parabolic-hyperbolic form is also affected as these equations contain tangential derivatives of $\kkappa$. Nevertheless, provided that $\mathcal{R}>-\tfrac12$ holds, the obtained modified parabolic-hyperbolic system is guaranteed to be well posed, and, more importantly, the initial data that was yielded by this method was also found to be strongly asymptotically flat. The viability of this proposal of Beyer {\it et al.}\,\cite{Beyer:2020zlo} was confirmed in our follow-up paper \cite{Csukas:2019qco}, where we also proposed altering part of the freely specifiable data in case of the algebraic-hyperbolic system. It is fair to admit, however, that the success of the latter proposal was limited.

	Several questions remained open for further study even after this significant progress reported in \cite{Beyer:2020zlo}. For example, are there other suitable modifications of the free part of the data? Are there modifications that do not affect the principal parts of the evolutionary forms of the constraints? Could the more general nonstrictly Kerr initial data configurations be asymptotically flat? Is it possible to control the rate of decay of the initial data to produce weakly or strongly asymptotically flat solutions?

	The results reported in this paper were inspired by curiosity about the above questions. Our motivation also benefited from a recent work by Beyer and Ritchie \cite{Beyer:2021kmi} demonstrating that the original parabolic-hyperbolic system---without applying any alteration---is suitable to yield asymptotically hyperboloidal initial datasets. Analogous support was provided by the success of constructing initial data---even though these are relevant only for the strong field regime---using the parabolic-hyperbolic formulation of the constraints, for single or binary rotating Kerr black hole configurations in \cite{Nakonieczna:2017eev, Doulis:2019egi}.

	The objectives set out in the present paper are many and varied. For the first time, we investigate the asymptotic behavior of the space of initial datasets in a small neighborhood of the initial data obtained on a Kerr-Schild time slice of a rotating Kerr black hole. We do this by modifying only the monopole part of some freely specifiable variables. This, on the one hand, guarantees that no alteration of the principal parts of the evolutionary forms of the constraints will occur. Hence the well posedness of the relevant partial differential equation (PDE) systems is expected to hold, based on the results covered in \cite{Racz:2015mfa}. On the other hand, while the method proposed in \cite{Beyer:2020zlo} applies only to the parabolic-hyperbolic system,  our proposal can also be used in solving the algebraic-hyperbolic system by controlling the monopole part of the freely specifiable secondary lapse, $\Nh$. In addition, we also introduce a straightforward method, applicable to both of the evolutionary forms of the constraints, that allows us to control the falloff rate of the monopole part of the constrained variable $\KK$, thereby generating weakly and strongly asymptotically flat initial data with any desired falloff rate. The viability of the introduced new methods is verified by integrating the evolutionary forms of the constraint equations numerically in the case of various near-Kerr initial data configurations.

	The structure of the paper is as follows: In Sec. \ref{sec:prelim}, we recall the notions and notations utilized in this paper. Section \ref{sec:swvariables} focuses on defining the most relevant spin-weighted variables, while Secs. \ref{sec:phformulation} and \ref{sec:ahformulation} recall the exact form of the equations and some of their properties concerning the parabolic-hyperbolic and algebraic-hyperbolic form of the constraints, respectively.
	In Sec. \ref{sec:coords} we select a suitable (Kerr-Schild) time slice foliation to the Kerr black hole background. As a closing to Sec. \ref{sec:prelim}, in Sec. \ref{sec:asym}, we recall the notion of strong and weak asymptotic flatness. Our main results are presented in Sec. \ref{sec:results}. Since in all the former studies, exclusively the asymptotics of near-Schwarzschild initial data was investigated, in Sec. \ref{sec:strictscheme}, we start by examining the asymptotics of strictly near-Kerr black hole initial data configurations. Although this does not yield the desired falloff behavior, we still find it beneficial to provide this for comparison and to verify that even in this more complicated case, it is indeed only the monopole part of $\KK$ that requires suitable control.
	Sections \ref{sec:phm} and \ref{sec:ahm} present our proposal applied to the parabolic-hyperbolic and the algebraic-hyperbolic formulations, respectively. These sections start with motivating our novel approach by inspecting the analytic solutions in spherical symmetry and then presenting the corresponding numerical results in the case of near-Kerr initial data configurations. We close this paper with our final remarks in Sec. \ref{sec:summary}.\,\footnote{Unless indicated otherwise, our conventions and notations are the same as in \cite{Wald:1984rg}. In particular, we use geometric units.}

  \section{Preliminaries}
  \label{sec:prelim}

  The geometric content of an initial dataset is represented by a pair of symmetric tensor fields $(h_{ab},K_{ab})$, where $h_{ab}$ is a Riemannian metric on a three-dimensional manifold $\Sigma$. Once $\Sigma$ is embedded into a four-dimensional spacetime $(M,g_{ab})$ the fields $h_{ab}$ and $K_{ab}$ get to be the induced metric and extrinsic curvature of $\Sigma$ in $(M,g_{ab})$. The fields $(h_{ab},K_{ab})$ are subject to the constraints which, in the vacuum case, read as
  \begin{align}
    \Rthree-K_{ab}K^{ab}+K^2&=0,\label{eq:Ham3}\\
    D_bK^b{}_a-D_aK&=0\label{eq:Mom3}\,,
  \end{align}
  where $D_a$ and $\Rthree$ are the covariant derivative operator and Ricci scalar  associated with $h_{ab}$, respectively, and $K$ denotes the trace of the extrinsic curvature.

  \subsection{Foliation based decompositions}
  \label{sec:geomvariables}

  The geometric construction underlying the evolutionary interpretation of the constraints lies on the assumption that the three-dimensional initial data surface, $\Sigma$, can be foliated by a one-parameter family of $2$-surfaces, $\mathcal{S}_\varrho$, which are the $\varrho=const$ level surfaces of a function $\varrho:\Sigma\rightarrow\mathbf{R}$ \cite{Racz:2014jra,Racz:2014gea,Racz:2015mfa}.

  The unit normal to the $\mathcal{S}_\varrho$ surfaces is $\widehat{n}_a=\Nh D_a\varrho$, such that $\widehat{n}^a\widehat{n}_a=1$. Then the induced metric on $\mathcal{S}_\varrho$ reads as
  \begin{equation}
    \widehat{\gamma}_{ab}=h_{ab}-\widehat{n}_a\widehat{n}_b\,,
  \end{equation}
  whereas the operator projecting fields defined on $\Sigma$ onto $\mathcal{S}_\varrho$ is $\widehat{\gamma}^a{}_b=h^{ae}{\gamma}_{eb}$. The covariant derivative operator associated with $\widehat{\gamma}_{ab}$ is denoted by $\widehat{D}_a$, whereas the corresponding Ricci scalar and the extrinsic curvature of the $2$-surfaces by $\Rh$ and $\Kh_{ab}$, respectively. To uncover implicit involvements of the lapse, $\Kh_{ab}$ will be replaced by the product $\Nh^{-1}\Kstar_{ab}$, where $\Kstar_{ab}=\tfrac12\,\mathscr{L}_{[\rho-\Nh]}\widehat{\gamma}_{ab}$.

  A radial flow vector field, $\rho^a$, is also chosen such that its integral curves  intersect each $\mathcal{S}_\varrho$ level surface precisely once, and it is normalized such that $\rho^aD_a\varrho=1$. Then the lapse and shift associated with $\rho^a$ are $\Nh=\rho^a\widehat{n}_a$ and $\Nh^a=\widehat{\gamma}^a{}_b\rho^b$, respectively. Using these variables the geometric content of the metric, $h_{ab}$, can be uniquely represented by the triplet $(\Nh,\Nh^a,\widehat{\gamma}_{ab})$ \cite{Csukas:2019qco}.

  The extrinsic curvature, $K_{ab}$, can also be decomposed as
  \begin{equation}
    K_{ab}=\kkappa\,\widehat{n}_a\widehat{n}_b+\kk_a\widehat{n}_b+\kk_b\widehat{n}_a+\KK_{ab},
  \end{equation}
  where
  \begin{equation}\label{eq:decompKK}
  \kkappa=K_{ab}\,\widehat{n}^a\widehat{n}^b\,, \quad \kk_a=K_{bc}\,\widehat{\gamma}^b{}_a\widehat{n}^c\,, \quad \KK_{ab}=K_{cd}\,\widehat{\gamma}^c{}_a\widehat{\gamma}^d{}_b\,.
  \end{equation}
   The tensorial projection of the extrinsic curvature can further be split into its trace and trace-free parts, which are given as
  \begin{equation}\label{eq:traceKK}
   \KK=\KK_{ab}\widehat{\gamma}^{ab}  \quad {\rm and}  \quad \Kc_{ab}=\KK_{ab}-\tfrac{1}{2}\,\KK\,\widehat{\gamma}_{ab}\,.
  \end{equation}
  Note finally that the septet $(\Nh,\Nh^a,\widehat{\gamma}_{ab};\kkappa,\kk_a,\KK,\Kc_{ab})$ is algebraically equivalent to the geometric content represented by the pair $(h_{ab}, K_{ab})$ on $\Sigma$.

	\subsection{Spin-weighted variables}
	\label{sec:swvariables}

	As argued in \cite{Racz:2015ena, Racz:2016wcs, Racz:2017krc}, it is rewarding to use spin-weighted variables. In doing so, one first introduces a complex null dyad, $\{q_a,\overline{q}_a\}$, where bar denotes complex conjugate, on one of the $\mathcal{S}_{\varrho}$ level surfaces, say on $\mathcal{S}_{\varrho_0}$ such that $q_{ab}=q_a\overline{q}_b$ is required to be the unit sphere metric on $\mathcal{S}_{\varrho_0}$. Then the dyad $\{q_a,\overline{q}_a\}$ is automatically normalized as $q^{ab}q_a\overline{q}_b=2$, whereas the indices of the dyad are raised and lowered by $q^{ab}$ and $q_{ab}$, respectively. To get a complex null dyad, $\{q_a,\overline{q}_a\}$ on each of the $\varrho=const$ level surfaces the dyad fixed on $\mathcal{S}_{\varrho_0}$ is Lie-dragged with respect to $\rho^a$ onto $\Sigma$. Note that then the action of the tangential derivative $\widehat{D}_a$ can be rephrased using the Newman-Penrose $\eth$ and $\overline{\eth}$ operators \cite{Racz:2017krc}. The basic variables we use in the succeeding subsections, including the spin-weighted ones, are listed in Table\,\ref{table:variables}.
	\begin{table}[H]
		\centering  \hskip-.15cm
		\begin{tabular}{|c|c|c|}
			\hline Notation &  Definition  & \hskip-0.7cm$\phantom{\frac{\frac12}{A}_{B_D}}$
			Spin-weight \\ \hline \hline

			$\mathbf{a}$ &  $\tfrac12\,q^i\,\overline q^j\,\widehat\gamma_{ij}$  & \hskip-0.7cm$
			\phantom{\frac{\frac12}{A}_{B_D}}$ $0$ \\  \hline

			$\mathbf{b}$ &  $\tfrac12\,q^i q^j\,\widehat\gamma_{ij}$
			& \hskip-0.7cm$\phantom{\frac{\frac12}{A}_{B_D}}$ $2$ \\  \hline

			$\mathbf{d}$ &  $\mathbf{a}^2-\mathbf{b}\,\overline{\mathbf{b}}$
			& \hskip-0.7cm$\phantom{\frac{\frac12}{A}_{B_D}}$ $0$ \\  \hline

			$\mathbf{A}$ &  $q^a q^b {C^e}{}_{ab}\,\overline q_e
			= \mathbf{d}^{-1}\left\{ \mathbf{a}\left[2\,\eth\,\mathbf{a}
			-\,\overline{\eth}\,\mathbf{b}\right]
			-  \,\overline{\mathbf{b}}\,\eth\,\mathbf{b} \right\} $
			& \hskip-0.7cm$\phantom{\frac{\frac12}{A}_{B_D}}$ $1$ \\  \hline

			$\mathbf{B}$ &  $\,\overline q^a q^b {C^e}{}_{ab}\,q_e
			= \mathbf{d}^{-1}\left\{ \mathbf{a}\,\overline{\eth}\,\mathbf{b}
			- \mathbf{b}  \,\eth\,\overline{\mathbf{b}}\right\}$
			& \hskip-0.7cm$\phantom{\frac{\frac12}{A}_{B_D}}$ $1$ \\  \hline

			$\mathbf{C}$ &  $q^a q^b {C^e}{}_{ab}\,q_e
			= \mathbf{d}^{-1}\left\{ \mathbf{a}\,\eth\,\mathbf{b}
			-  \mathbf{b}\left[2\,\eth\,\mathbf{a}
			-\,\overline{\eth}\,\mathbf{b}\right] \right\}$
			& \hskip-0.7cm$\phantom{\frac{\frac12}{A}_{B_D}}$ $3$ \\  \hline

			$\,\widehat{R}$ &  $\tfrac12\, {\mathbf{a}}^{-1}\left(2\, \mathbf{R}
			- \left\{ \,  \eth\,\overline{\mathbf{B}} - \overline{\eth}\,\mathbf{A}
			- \tfrac12\,\left[\, \mathbf{C}\,\overline{\mathbf{C}}
			- \mathbf{B}\,\overline{\mathbf{B}} \,\right]\, \right\}\,\right)$
			& \hskip-0.7cm$\phantom{\frac{\frac12}{A}_{B_D}}$ $0$ \\  \hline

			${\mathbf{N}}$ &  $q_i\widehat N^i$
			& \hskip-0.7cm$\phantom{\frac{\frac12}{A}_{B_D}}$ $1$ \\  \hline

			$\mathbf{k}$ &  $q^i {\rm\bf k}{}_{i}$  &
			\hskip-0.7cm$\phantom{\frac{\frac12}{A}_{B_D}}$ $1$
			\\  \hline

			$\mathbf{K}$ &  $ \widehat\gamma^{kl} \,{\rm\bf K}{}_{kl}
			$  & \hskip-0.7cm$\phantom{\frac{\frac12}{A}_{B_D}}$ $0$ \\  \hline

			$\Kc{}_{qq}$ &  $q^kq^l\,\Kc{}_{kl}
			$  & \hskip-0.7cm$\phantom{\frac{\frac12}{A}_{B_D}}$ $2$ \\  \hline

			$\Kc{}_{q\overline{q}}$ &  $q^k\,\overline{q}^l\,\Kc{}_{kl}
			= (2\,\mathbf{a})^{-1} [\,\mathbf{b}\,\overline{\Kc{}_{qq}}
			+  \overline{\mathbf{b}}\,\Kc{}_{qq} \,]
			$  & \hskip-0.7cm$\phantom{\frac{\frac12}{A}_{B_D}}$ $0$ \\  \hline

			$\,\Kstar$ &  ${\Kstar} = \widehat\gamma^{ij} \Kstar_{ij}  $
			& \hskip-0.7cm$\phantom{\frac{\frac12}{A}_{B_D}}$ $0$ \\  \hline

			$\Kstar{}_{qq}$ &  $q^i q^j\Kstar_{ij}
			= \tfrac12\,\left\{2\,\partial_\rho\mathbf{b} - 2\,\eth\,\mathbf{N}
			+ {\mathbf{C}}\,\overline {\mathbf{N}} +\mathbf{A} \,{\mathbf{N}} \,  \right\} $
			& \hskip-0.7cm$\phantom{\frac{\frac12}{A}_{B_D}}$ $2$ \\  \hline

			$\Kstar{}_{q\overline{q}}$ &  $q^k\,\overline{q}^l\,\Kstar {}_{kl}
			=  {\mathbf{a}}^{-1}\{\,\mathbf{d}\cdot\Kstar\}
			+ \tfrac12 \,[\,\mathbf{b}\,\overline{\Kstar}{}_{qq}
			+  \overline{\mathbf{b}}\,\Kstar{}_{qq}  \,]\,\}
			$  & \hskip-0.7cm$\phantom{\frac{\frac12}{A}_{B_D}}$ $0$ \\  \hline

		\end{tabular}
		\caption{\small The variables applied in providing the evolutionary form of the constraints.}\label{table:variables}
	\end{table}
  Note that, in virtue of $\Kc_{ab}\widehat{\gamma}^{ab}=0$, for the contraction $\Kcqqb=\Kc_{ab}q^a\overline{q}^b$, as indicated in Table\,\ref{table:variables},
  \begin{equation}
  	\Kcqqb=(2\aaa)^{-1}\big[\bb\overline{\Kcqq}+\overline{\bb}\Kcqq\big]
  \end{equation}
  holds. Note also that $\widehat{\gamma}_{ab}$ can be given, in terms of the dyad $\{q_a,\overline{q}_a\}$, and the variables $\aaa$ and $\bb$, as
  \begin{equation}
    \label{eq:gammahat}
  	\widehat{\gamma}_{ab}=\aaa\, q_{ab}+\tfrac{1}{2}[\bb\,\overline{q}_a\overline{q}_b+\overline{\bb}\,q_aq_b]\,.
  \end{equation}

  Note, finally, that the geometric content of $(h_{ab}, K_{ab})$ can also be represented by the octet $(\Nh,\NNt,\aaa,\bb;\kkappa,\kk,\KK,\Kcqq)$. As shown in \cite{Racz:2015mfa,Racz:2015ena,Racz:2016wcs} by choosing $\Nh$, $\kk$, and $\KK$ as constrained variables, we arrive at the parabolic-hyperbolic formulation while selecting $\kkappa$, $\kk$, and $\KK$ as constrained variables we arrive at the algebraic-hyperbolic formulation of the constraints.

  \subsection{The parabolic-hyperbolic equations}
  \label{sec:phformulation}

 Selecting $\Nh$, $\kk$, $\KK$ as constrained variables, the Hamiltonian and momentum constraints read as \cite{Racz:2017krc}
  \begin{multline}
      \label{eq:phN}
      \Kstar\left[\partial_r\Nh-\tfrac12\,\Nt\,\ethb\Nh-\tfrac12\,\Ntb\,\eth\Nh\right]\\
      -\tfrac12\,\dd^{-1}\Nh^2\left[\,\aaa\left\{\eth\ethb\Nh-\BB\,\ethb\Nh\right\}-
      \bb\left\{\ethb^2\!\Nh-\tfrac12\,\AAAb\,\ethb\Nh-\tfrac12\,\CCb\,\eth\Nh\right\}+cc.\right]\\
      -\mathcal{A}\,\Nh-\mathcal{B}\,\Nh^{\,3}=0\,,
  \end{multline}
  \begin{equation}
      \label{eq:phk}
      \partial_r\kk-\tfrac12\,\Nt\,\ethb\kk-\tfrac12\,\Ntb\,\eth\kk-\tfrac12\,\Nh\,\eth\KK+\ff_{PH}=0\,,
  \end{equation}
  \begin{equation}
      \label{eq:phK}
      \partial_r\KK-\tfrac12\,\Nt\,\ethb\KK-\tfrac12\,\Ntb\,\eth\KK-
      \tfrac{1}{2}\,\Nh\,\dd^{-1}\Big\{\,\aaa(\eth\kkb+\ethb\kk)-\bb\,\ethb\kkb-\bbb\,\eth\kk\,\Big\}+\FF_{PH}=0\,,
  \end{equation}
  where the coefficients $\mathcal{A}$, $\mathcal{B}$, and the source terms $\ff_{PH}$, $\FF_{PH}$ in \eqref{eq:phN}, \eqref{eq:phk}, and \eqref{eq:phK}, are given as
  \begin{equation}\label{eq:phA}
  	\mathcal{A}=\partial_r\Kstar-\tfrac12\,\Nt\,\ethb\Kstar-\tfrac12\Ntb\,\eth\Kstar+\tfrac12\,\Big[\Kstar{}^2+\Kstar_{kl}\Kstar{}^{kl}\Big]\,,
  \end{equation}
  \begin{equation}\label{eq:phB}
  	\mathcal{B}=-\tfrac12\,\Big[\Rh+2\,\kkappa\,\KK+\tfrac12\,\KK^2-\dd^{-1}[2\,\aaa\,\kk\,\kkb-\bb\,\kkb^2-\bbb\,\kk^2]-\Kc{}_{kl}\Kc{}^{kl}\Big]\,,
  \end{equation}
  \begin{multline}\label{eq:phf}
    \ff_{PH}=-\tfrac12\,\Big[\kk\,\eth\Ntb+\kkb\,\eth\Nt\Big]-\left[\kkappa-\tfrac12\,\KK\right]\eth\Nh+\Kstar\,\kk-\Nh\Big[\eth\kkappa+q^i\indot{\widehat{n}}{}^l\Kc{}_{li}-q^i\widehat{D}^l\Kc{}_{li}\Big]\,,
  \end{multline}
  \begin{align}\label{eq:phF}
    \FF_{PH}=\tfrac{1}{4}\Nh\,\dd^{-1}\Big\{2\,\aaa\,\BB\,\kkb-\bb(\,\CCb\,\kk+\AAAb\,\kkb)+cc.\Big\}
    & -\dd^{-1}\Big[(\aaa\,\kkb-\bbb\,\kk)\,\eth\Nh+cc.\Big]\nonumber \\ & +\Big[\Kc_{ij}{\Kstar}{}^{ij}-\left(\kkappa-\tfrac12\,\KK\right)\Kstar\Big]\,
  \end{align}
  with $cc.$ denoting the complex conjugate of the preceding terms, while the explicit form of the terms, such as $\Kstar_{ij}{\Kstar}{}^{ij}$, $\Kc_{ij}{\Kstar}{}^{ij}$, $\Kc_{ij}{\Kc}{}^{ij}$, $q^i\indot{\widehat{n}}{}^l\Kc{}_{li}$, $q^i\widehat{D}^l\Kc{}_{li}$, can be found in \cite{Racz:2017krc}.

  As shown in \cite{Racz:2015mfa}, \eqref{eq:phN} is a Bernoulli-type parabolic equation, while \eqref{eq:phk} and \eqref{eq:phK} form a first-order symmetric hyperbolic system. This coupled parabolic-hyperbolic system always gets to be well posed in those subregions of $\Sigma$ where the positivity of $\Kstar$ can be guaranteed. As $\Kstar$ is determined by the freely specifiable variables $\aaa$, $\bb$, and $\NNt$, the coupled parabolic-hyperbolic system can always be guaranteed to be well posed at least on a one-sided neighborhood of the initial data surface  $\mathcal{S}_{\varrho_0}$\,\footnote{The initial data for the evolutionary form of the constraints, specified on one of the ${\varrho}=const$ level sets in $\Sigma$ should not be confused with an initial data set $(\Sigma; h_{ab}, K_{ab})$ relevant for the time-evolution part of the Einstein's equations.} in  $\Sigma$. Note also that besides the freely specifiable eight real functions, represented by the variables $(\NNt,\aaa,\bb;\kkappa,\Kcqq)$, on  $\Sigma$, we also have the freedom to choose initial data $(\Nh|_{\varrho_0},\kk|_{\varrho_0},\KK|_{\varrho_0})$ to the parabolic-hyperbolic equations \eqref{eq:phN}-\eqref{eq:phK} on $\mathcal{S}_{\varrho_0}$.

  \subsection{The algebraic-hyperbolic equations}
  \label{sec:ahformulation}

  Choosing $\KK$, $\kk$, and $\kkappa$ as basic variables the constraints read as \cite{Racz:2017krc}
  \begin{equation}
      \label{eq:ahK}
      \partial_r\KK-\tfrac12\,\Nt\,\ethb\KK-\tfrac12\,\Ntb\,\eth\KK-\tfrac{1}{2}\,\Nh\,\dd^{-1}\Big\{\aaa(\eth\kkb+\ethb\kk)-\bb\,\ethb\kkb-\bbb\,\eth\kk\Big\}+\FF_{AH}=0\,,
  \end{equation}
  \begin{multline}
      \label{eq:ahk}
      \partial_r\kk-\tfrac12\,\Nt\,\ethb\kk-\tfrac12\,\Ntb\,\eth\kk+\Nh\,\KK^{-1}\Big\{\kkappa\,\eth\KK-\dd^{-1}\big[(\aaa\kk-\bb\kkb)\,\eth\kkb
      +(\aaa\kkb-\bbb\kk)\,\eth\kk\big]\Big\}+\ff_{AH}=0\,,
  \end{multline}
	\begin{equation}
		\label{eq:ahkappa}
			\kkappa=\tfrac{1}{2}\KK^{-1}\Big[\dd^{-1}\big(2\aaa\,\kk\kkb-\bb\,\kkb^2-\bbb\,\kk^2\big)-\tfrac12\KK^2-\kkappa_0\Big]\,,
	\end{equation}
with
	\begin{equation}
		\kkappa_0=\Rthree-\Kc_{kl}\Kc{}^{kl}\,, \label{eq:ahkappa0}
	\end{equation}
  where the pertinent source terms $\FF_{AH}$, $\ff_{AH}$ are given as
  \begin{align}\label{eq:ahF}
      \FF_{AH}=  \tfrac{1}{4}\,\Nh\,\dd^{-1}\Big\{2\,\aaa\,\BB\,\kkb-\bb(\,\CCb\,\kk+\AAAb\,\kkb)+cc.\Big\}
      & -\dd^{-1}\Big[(\aaa\kkb-\bbb\kk)\,\eth\Nh+cc.\Big] \nonumber \\ & +\Big[\Kc_{ij}{\Kstar}{}^{ij}-\left(\kkappa-\tfrac12\,\KK\right)\Kstar\Big]\,,
  \end{align}
  \begin{align}\label{eq:ahf}
      \ff_{AH} = &-\tfrac12\,\Big[\kk\,\eth \Ntb +\kkb\,\eth \Nt\Big]\nonumber \\
        &+\tfrac{1}{2}\,\Nh\,(\dd\cdot\KK)^{-1}\Big[(\aaa\,\kk-\bb\,\kkb)(\BBb\,\kk+\BB\,\kkb)+(\aaa\,\kkb-\bbb\,\kk)(\CC\,\kkb+\AAA\,\kk)\Big] \nonumber \\ &
      -\left[\kkappa-\tfrac12\,\KK\right]\eth\Nh+\Nh\Big[\tfrac{1}{2}\KK^{-1}\eth\kkappa_0+\Nh^{-1}\Kstar\,\kk-q^i\indot{\widehat{n}}{}{}^l\Kc_{li}+q^i\widehat{D}^l\Kc_{li}\Big]\,.
  \end{align}

	As shown in \cite{Racz:2015mfa}, \eqref{eq:ahK}-\eqref{eq:ahk} form a first-order Friedrichs symmetrizable hyperbolic system for the vector valued variable $(\KK,\kk)$ provided that the inequality $\kkappa\cdot\KK<0$ holds. Notice that as this latter condition refers to the dependent variables, its validity has to be checked during the integration of the underlying system. Note also that in doing so the fields $(\Nh,\NNt,\aaa,\bb;\Kcqq)$ can be specified at will throughout $\Sigma$, along with  the Cauchy data for the system \eqref{eq:ahK}-\eqref{eq:ahk} constituted by the pair $(\KK|_{\varrho_0},\kk|_{\varrho_0})$ given on the $\varrho=\varrho_0$ initial data level set.

  \subsection{Foliations of Kerr-Schild time slices}
  \label{sec:coords}

	As indicated in the Introduction, one of our principal aims is to explore the asymptotic behavior of at least a small neighborhood of the initial datasets deduced from a rotating Kerr black hole on a Kerr-Schild time slice. Accordingly, such as in all the former investigations  \cite{Racz:2015ena, Beyer:2017njj, Beyer:2019kty, Beyer:2020zlo, Beyer:2021kmi, Csukas:2019qco}, we start by deducing a distinguished initial dataset, represented by $(\Nh,\NNt,\aaa,\bb;\kkappa,\kk,\KK,\Kcqq)$, relevant for a Kerr spacetime, with mass and rotation parameters $M$ and $a$, respectively, on a $t_{KS}=const$ Kerr-Schild time slice. These time slices are preferable as they connect the black hole interior with spacelike infinity. 

    The Kerr-Schild form
    \begin{equation}\label{eq:KSM}
    	g_{ab}=\eta_{ab}+2 H l_al_b\,,
    \end{equation}
	with
	\begin{equation}
		H=\frac{M r^3}{r^4+ a^2 \tilde{z}^2}\,, \quad l_a=\left(1,\frac{r\tilde{x} + a 	\tilde{y}}{r^2+a^2},\frac{r\tilde{y} - a \tilde{x}}{r^2+a^2},\frac{\tilde{z}}{r}\right)\,,
	\end{equation}
  is probably one of the simplest and most widely used representation of the Kerr black hole spacetime, with spin pointing to the positive $\tilde{z}$ direction. In  \eqref{eq:KSM} $\eta_{ab}$ stands for the metric of an auxiliary Minkowski spacetime with inertial coordinates  $(\tilde{t},\tilde{x},\tilde{y},\tilde{z})$. The spatial part of these coordinates $(\tilde{x},\tilde{y},\tilde{z})$ and the Boyer-Lindquist radial coordinate, $r$, are related via the implicit relation
  \begin{equation}\label{eq:BL-coords}
  	\frac{\tilde{x}^2+\tilde{y}^2}{r^2+a^2}+\frac{\tilde{z}^2}{r^2}=1\,.
  \end{equation}
	The $t_{KS}=const$ Kerr-Schild time slices of a Kerr spacetime are nothing but $\tilde{t}=const$ time-level surfaces of the auxiliary Minkowski spacetime. As  $(\tilde{x},\tilde{y},\tilde{z})$ are Cartesian coordinates on the $\tilde{t}=const$ time-level surfaces, it is tempting to foliate the Kerr-Schild time slices by the $\tilde{x}^2+\tilde{y}^2+\tilde{z}^2=const$ level sets. Nevertheless, however controversial it may sound, the coordinates $(\tilde{x},\tilde{y},\tilde{z})$ do not form a suitable admissible asymptotic system for the Kerr solution, which (as we shall see in the following subsection) plays a central role in setting up the precise formulation of asymptotic flatness of the data specified on $t_{KS}=const$ Kerr-Schild time slices.

	For this purpose the spherical Kerr-Schild coordinates, introduced by Chen {\it et al.}\,\cite{Chen:2021rtb}, turned out to be more appropriate. The spherical Kerr-Schild coordinates, $(x,y,z)$, are related to $(\tilde{x},\tilde{y},\tilde{z})$ via the relations
  \begin{equation}
    \label{eq:sphKS}
    x=\frac{r\tilde{x}}{\sqrt{r^2+a^2}},\qquad y=\frac{r\tilde{y}}{\sqrt{r^2+a^2}},\qquad z=\tilde{z}\,.
  \end{equation}
	As noted in \cite{Chen:2021rtb}, while in the original Kerr-Schild coordinates  $(\tilde{x},\tilde{y},\tilde{z})$, and with respect to the Euclidean metric, $\delta_{ab}$,  the Boyer-Lindquist $r=const$ level sets are, in virtue of \eqref{eq:BL-coords}, oblate spheroids, the same level sets, at least in an affine sense, are the $x^2+y^2+z^2=r^2$ spheres with respect to the spherical Kerr-Schild coordinates. Note that the $\tilde{x}^2+\tilde{y}^2+\tilde{z}^2=const$ and $x^2+y^2+z^2=const$ level sets do coincide in the $a\rightarrow0$ Schwarzschild limit.

	The payback of using the spherical Kerr-Schild coordinates $(x,y,z)$ is that they form admissible coordinates on the $t_{KS}=const$ Kerr-Schild time slices. This gets more transparent when one is applying the spherical coordinates $(r,\vartheta,\varphi)$  defined via the (standard) implicit relations\,\footnote{It is important to emphasize that the $(\tilde{x},\tilde{y},\tilde{z})$ oblate spheroidal coordinates can be given in terms of the Boyer-Lindquist coordinates $(r,\theta,\phi)$ as
	\begin{equation}\label{eq:oblatexyz}
		\tilde{x}=\sqrt{r^2+a^2}\,\cos\phi\,\sin\theta\,,\quad \tilde{y}=\sqrt{r^2+a^2}\,\sin\phi\,\sin\theta\,,\quad \tilde{z}=r\,\cos\theta\,.
	\end{equation}
	Note that the Boyer-Lindquist angular coordinates $(\theta,\phi)$ and those in \eqref{eq:sphericalxyz} are related as $\theta=\vartheta$ and $\phi=\varphi-\arctan(a/r)$. It is indeed the very last relation that explains why the $(x,y,z)$ spherical Kerr-Schild coordinates could be suitable, and, in contrast, the $(\tilde{x},\tilde{y},\tilde{z})$ oblate spheroidal coordinates cannot form asymptotically admissible coordinates on Kerr-Schild time slices.}
  \begin{equation}\label{eq:sphericalxyz}
    x=r\,\cos\varphi\,\sin\vartheta\,,\quad y=r\,\sin\varphi\,\sin\vartheta\,,\quad z=r\,\cos\vartheta\,.
  \end{equation}
	Utilizing these spherical coordinates $(r,\vartheta,\varphi)$ the data $(\Nh,\NNt,\aaa,\bb;\kkappa,\kk,\KK,\Kcqq)$, relevant for the considered Kerr spacetime on $\Sigma$, i.e., on the $t=0$ Kerr-Schild time slice,  foliated by the $r=const$ level sets and with respect to the flow $\rho^a=(\partial_r)^a$, can be given as
  \begin{equation}
	  \label{eq:kerrNh-kerrNN}
	  \Nh=\sqrt{\frac{(1+2H)\Upsilon}{\Xi}}\,, \qquad \NN=\imath\,\frac{2aMr\sin\vartheta}{(a^2+r^2)\,\Xi}\,,
  \end{equation}
  \begin{equation}
	  \label{eq:kerraa-kerrbb}
	  \aaa=a^2+r^2-\tfrac{1}{2}a^2(1-2H)\sin^2\vartheta\,, \qquad\bb=-\tfrac{1}{2}a^2(1+2H)\sin^2\vartheta\,,
  \end{equation}
  \begin{multline}
	  \label{eq:kerrkappa}
	  \kkappa=\frac{2a^2\sqrt{1+2H}\sin^2\vartheta\, \partial_r H}{\Upsilon}-\frac{2\,(1+H)\,\Xi\, \partial_r H}{\Upsilon\,(1+2H)^{3/2}}\\
	  -\frac{2a^2H\sqrt{1+2H}\sin^2\vartheta\,(r+a^2\sin^2\vartheta\, \partial_r H)}{\Upsilon\,\Xi}\,,
  \end{multline}
  \begin{multline}
	  \label{eq:kerrkk}
	  \kk=-\frac{a^2\sin^2\vartheta\, \partial_\vartheta H}{\sqrt{\Upsilon\,\Xi}}+\sqrt{\frac{\Xi}{\Upsilon}}\,\frac{\partial_\vartheta H}{1+2H}\\
	  -\imath\left(a\sin\vartheta\,\sqrt{\frac{\Xi}{\Upsilon}}\,\partial_r H-\frac{2aH\sin\vartheta\,(r+a^2\sin^2\vartheta\, \partial_r H)}{\sqrt{\Upsilon\,\Xi}}\right)\,,
  \end{multline}
  \begin{equation}
	  \label{eq:kerrKK}
	  \KK=-\frac{H-r\, \partial_r H}{r\sqrt{1+2H}}-\frac{2H(r+a^2\sin^2\vartheta\, \partial_r H)}{\sqrt{1+2H}\,\Xi}\,,
  \end{equation}
  \begin{equation}
	  \label{eq:kerrKcqq}
	  \Kcqq=\frac{\Upsilon+\Xi}{\sqrt{1+2H}}\left(\frac{H(r+a^2\sin^2\vartheta\, \partial_r H)}{\Xi}-\frac{H-r \,\partial_r H}{2r}\right)+\imath\,\frac{2a\sin\vartheta\, \partial_\vartheta H}{\sqrt{1+2H}}\,,
  \end{equation}
  where $\Upsilon=r^2 + a^2\cos^2\vartheta$, $H=Mr/\Upsilon$ and $\Xi=r^2+a^2+2a^2H\sin^2\vartheta$.

	Note that due to the symmetries of the Kerr spacetime, $\Nt$ is purely imaginary, and $\bb$ is real, although, in general, they are complex-valued functions. Note also that in the $a\rightarrow0$ Schwarzschild limit $\Nt$, $\bb$, $\kk$, and $\Kcqq$ all vanish, and that the variables $(\Nh,\NNt,\aaa,\bb;\kkappa,\kk,\KK,\Kcqq)$ reduce to the formulas applied in \cite{Csukas:2019qco}.

	To avoid potential numerical inaccuracies in evaluating some of the more complex geometric quantities, for instance, $\Rthree$, it turned out to be rewarding to inspect if subterms of the same magnitude with opposite signs are involved. In practice, it was profitable to simplify the corresponding expressions using symbolic computer algebra and evaluate the yielded formulas numerically, in accordance with similar experiences reported in \cite{Beyer:2017njj}.

  \subsection{Asymptotic flatness}
  \label{sec:asym}

	An initial dataset $(\Sigma, h_{ab}, K_{ab})$ is considered to be strongly asymptotically flat if the complement of a compact set in $\Sigma$ can be mapped by an admissible coordinate system $(x,y,z)$ diffeomorphically onto the complement of a closed ball in $\mathbf{R}^3$ such that in these coordinates
	\begin{align}
    \label{eq:strongasymhK}
		h_{ab}&=\left(1+\tfrac{C}{r}\right)\delta_{\alpha\beta}+\mathcal{O}(r^{-2}),\\
		K_{ab}&\sim\mathcal{O}(r^{-2})
	\end{align}
	hold as $r=\sqrt{x^2+y^2+y^2}$ goes to infinity \cite{Dain:2001ry}, where $\delta_{ab}$ denotes the components of the flat three-metric in the admissible Cartesian coordinates $(x,y,z)$. These conditions guarantee that the ADM mass, momentum, and angular momentum of the initial dataset are well defined.

	A weaker notion of asymptotic flatness is also frequently used. An initial dataset $(\Sigma, h_{ab}, K_{ab})$ will be referred to as weakly asymptotically flat if the milder falloff conditions
	\begin{align}
    \label{eq:weakasymhK}
		h_{ab}&=\left(1+\tfrac{C}{r}\right)\delta_{ab}+\mathcal{O}(r^{-3/2-\epsilon}),\\
		K_{ab}&\sim\mathcal{O}(r^{-3/2-\epsilon})
	\end{align}
	hold, for some small positive number $\epsilon$. Note that these weaker conditions guarantee that the ADM mass and linear momentum are well defined \cite{Chrusciel:1986xts}.

	It was shown in  \cite{Csukas:2019qco} that by transforming an admissible coordinate system $(x,z,y)$ to spherical coordinates $(r,\vartheta,\varphi)$, by applying the relation in \eqref{eq:sphericalxyz}, one can
  reformulate the falloff conditions \eqref{eq:strongasymhK} and \eqref{eq:weakasymhK} to those introduced in Sec. \ref{sec:geomvariables}. The corresponding falloff conditions for the metric components in the case of strong asymptotic flatness are \cite{Csukas:2019qco}
  \begin{equation}
    \Nh-1\sim\mathcal{O}(r^{-1})\,,\qquad\Nh^a\sim\mathcal{O}(r^{-3})\,,\qquad\widehat{\gamma}_{ab}\sim r^2q_{ab}+\mathcal{O}(r^0).
  \end{equation}

  Since the dyadic components are independent of $r$, the spin-weighted variables inherit the falloff conditions of the tensor quantities from which they are derived. Thus we have $\NN\sim\mathcal{O}(r^{-3})$, $\aaa\sim\mathcal{O}(r^2)$, and $\bb\sim\mathcal{O}(r^2)$. Looking at \eqref{eq:gammahat}, however, we can derive a stronger condition on $\bb$ by realizing that the part of $\widehat{\gamma}_{ab}$ proportional to $q_{ab}$ is entirely represented by $\aaa$, so $\bb$ must satisfy $\bb\sim\mathcal{O}(r^0)$.
  Similarly, in the case of strong asymptotic flatness, the components of the extrinsic curvature must satisfy the falloff conditions \cite{Csukas:2019qco},
  \begin{equation}
  	\Kc_{ab}\sim\mathcal{O}(r^0)\,,\qquad\KK\sim\mathcal{O}(r^{-2})\,,\qquad\kk_a\sim\mathcal{O}(r^{-1})\,,\qquad\kkappa\sim\mathcal{O}(r^{-2}),
  \end{equation}
  which results in $\Kcqq\sim\mathcal{O}(r^0)$ and $\kk\sim\mathcal{O}(r^{-1})$.

  The background variables \eqref{eq:kerrNh-kerrNN}-\eqref{eq:kerrKcqq} have the asymptotic behavior
  \begin{align}
    \Nh&\sim1+\tfrac{M}{r}+\mathcal{O}(r^{-2})\,,&\NN&\sim\tfrac{2\imath a M\sin\vartheta}{r^3}+\mathcal{O}(r^{-4})\,,\\
    \aaa&\sim r^2+\mathcal{O}(r^0)\,,&\bb&\sim-\tfrac{1}{2}a^2\sin^2\vartheta+\mathcal{O}(r^{-1})\,,\\
    \kkappa&\sim\tfrac{2M}{r^2}+\mathcal{O}(r^{-3})\,,&\kk&\sim\tfrac{3\imath a M\sin\vartheta}{r^2}+\mathcal{O}(r^{-3})\,,\\
    \KK&\sim-\tfrac{4M}{r^2}+\mathcal{O}(r^{-3})\,,&\Kcqq&\sim-\tfrac{2a^2M\sin^2\vartheta}{r^2}+\mathcal{O}(r^{-3})
  \end{align}
  which correspond to the derived falloff conditions.

  Since this work is concerned by their asymptotic behavior, we recall that  the falloff properties of the constrained spin-weighted variables read as
	\begin{equation}\label{eq:stro-asymp}
		\Nh-1\sim\mathcal{O}(r^{-1}),\qquad\kkappa\sim\mathcal{O}(r^{-2}),\qquad\kk\sim\mathcal{O}(r^{-1}),\qquad\KK\sim\mathcal{O}(r^{-2})
	\end{equation}
	for strongly asymptotically flat initial data, while as
	\begin{equation}\label{eq:weak-asymp}
		\Nh-1\sim\mathcal{O}(r^{-1}),\quad\kkappa\sim\mathcal{O}(r^{-3/2-\epsilon}),\quad\kk\sim\mathcal{O}(r^{-1/2-\epsilon}),\quad\KK\sim\mathcal{O}(r^{-3/2-\epsilon})
	\end{equation}
	for weakly asymptotically flat data.

  \section{The main results: applying the new method}
  \label{sec:results}

  This section introduces the new method that allows us to control the falloff properties of the solutions to the evolutionary forms of the constraint equations. In doing so, all the variables are expanded using spin-weighted spherical harmonics, as done in \cite{Csukas:2019qco}. The replacement of a spin-weight $s$ variable ${}^{(s)}\mathbf{V}$ by the expansion
  \begin{equation}\label{eq: mult-exp}
  	{}^{(s)}\mathbf{V}(r,\vartheta,\varphi) = \sum_{\ell=|s|}^{\ell_{max}}\sum_{m=-\ell}^{\ell}\mathbf{V}_\ell{}^{m}(r)\cdot {}_s{Y_\ell}{}^m(\vartheta,\varphi)\,,
  \end{equation}
  where  ${}_s{Y_\ell}{}^m$ denotes the spin-weight $s$ spherical harmonics, allows us to evaluate its angular derivatives analytically as they can always be related to the $\eth$ and $\overline{\eth}$ operators.

  \subsection{The outline of the applied numerical scheme}
	\label{sec:numerical-scheme}

	In this study, we utilize the same numerical solver as applied in \cite{Csukas:2019qco}, which is based on spin-weighted spherical harmonics expansion in the angular sector and a fourth-order accurate adaptive Runge-Kutta-Fehlberg (RKF) method in solving the resulting ordinary differential equations for the expansion coefficients.\footnote{Although the code itself is not yet open source, its documentation is available to the public \cite{code}.} To ensure higher accuracy for our numerical solver, in each case, instead of solving \eqref{eq:phN}, \eqref{eq:phk} and \eqref{eq:ahK}-\eqref{eq:ahkappa}, we solve the equations which can be deduced from them for the nonlinear perturbations $\dNh=\Nh-\bgNh$, $\dKK=\KK-\bgKK$, and $\dkk=\kk-\bgkk$, in the parabolic-hyperbolic case, and for $\dKK$ and $\dkk$, in the algebraic-hyperbolic case. Here $\bgNh$, $\bgkk$, and $\bgKK$ signify the background quantities given by Eqs. \eqref{eq:kerrNh-kerrNN}, \eqref{eq:kerrkk}, and \eqref{eq:kerrKK}, respectively. The main payback of using the formulation based on deviations is that the corresponding dominant background fields do not hide the asymptotic behavior of the higher modes of the constraint fields. Note that the derivation of the equations governing the evolution of the nonlinear perturbations is straightforward. As they are slightly altered versions of the equations provided in Appendix A of \cite{Csukas:2019qco}, they will be omitted here.

	The input parameters in our numerical simulations are as follows: The integration of the deviation equations starts at $r=1$, that is, inside the black hole region. The non-negligible initial data there is
	\begin{equation}\label{eq:initial-data}
		\dKK|_{r=1}=-Y_2{}^0\,,
	\end{equation}
	which is the same order of magnitude as the dominant monopole part $\bgKK_0{}^0|_{r=1}=-5.7581$ of the background $\bgKK$ at $r=1$. As both the initial data in \eqref{eq:initial-data} and the background fields are axially symmetric, only the $m=0$ axisymmetric modes of the constrained fields get excited. Furthermore due to the parity symmetry of Kerr and the functional form of the dyad components, \eqref{eq:initial-data} excites only even $\ell$ modes of $\dKK$, $\dNh$, $\kkappa$, $\Re[\dkk]$, and only odd $\ell$ modes of $\Im[\dkk]$.

	The integration of the deviation equations goes out until reaching $r=10^9$, with a single exception (see Fig.\,\ref{fig:phbrm17} below), where it goes out to $r=10^{12}$. The mass and rotation parameters of the Kerr background take the values $M=1$ and $a=1/2$. Starting from the strong field region necessitates angular resolution with minimal cutoff $\ell_{max}=10$. This resolution allows us to compute the square roots with desired relative accuracy of $5\cdot10^{-18}$, which is the same accuracy as we demand in performing division by using Neumann series expansion. The error tolerance of RKF for each mode of a variable $f$ is set to $(f_\ell{}^m+\partial_rf_\ell{}^m)\epsilon$ with $\epsilon=10^{-5}$.

	Note also that the use of the \verb|WIGXJPF| library of Johansson and Forssén \cite{Johansson:2015cca} yielded notable improvements in the performance of our code. All the analytic calculations were derived or verified using {\it Mathematica} 12.3. In particular, the notebook we used to determine the background variables \eqref{eq:kerrNh-kerrNN}-\eqref{eq:kerrKcqq} (which serve as the main input to our code) is available as Supplemental Material.

  \subsection{Asymptotics of strictly near-Kerr initial datasets}
  \label{sec:strictscheme}

	As was mentioned in the Introduction, in the strictly near-Schwarzschild setup, neither the algebraic-hyperbolic formulation in the single Schwarzschild case \cite{Beyer:2017njj} nor the parabolic-hyperbolic formulation in the single and binary Schwarzschild black hole case \cite{Beyer:2019kty} allow suitable falloff for $\KK$.

	For instance, using the parabolic-hyperbolic formulation, Beyer {\it et al.}\,\cite{Beyer:2019kty} reported that besides that the lapse, $\Nh$, does not tend to the desired asymptotic value $1$, in general, the falloff rate of $\KK$ is $\mathcal{O}(r^{-1})$ that is also far too slow. In our follow-up mode-by-mode investigations, both of these observations were confirmed. Nevertheless, we also found that apart from the limiting value of $\Nh$ and the $\ell=0,m=0$ monopole mode of $\KK$, all the other modes of the involved variables falloff with a rate compatible with conditions in strong asymptotic flatness \cite{Csukas:2019qco}.

	Analogously, in the algebraic-hyperbolic case, Beyer {\it et al.}\,\cite{Beyer:2017njj} found that the falloff rate for $\KK$ is $\mathcal{O}(r^{-3/2})$ which is now strictly at the borderline not to allow weak asymptotic flatness. Surprisingly, the mode-by-mode investigations of similar strictly near-Schwarzschild initial datasets demonstrated that all the other modes of the involved variables falloff with a rate compatible with conditions in strong asymptotic flatness \cite{Csukas:2019qco}.

	Though in light of the results summarized above, one does not expect better asymptotic behavior of the constraint variables in case of strictly near-Kerr initial datasets, we devote this section to a short inspection of these configurations that still appears to be beneficial.

	It is rewarding to first glance at the asymptotic behavior of the background fields $\bgNh$, $\bgKK$, and $\bgkk$. Compatible with the strong asymptotic flatness of the Kerr solution, $\bgNh_0{}^0$ tends to the value $1$ with the rate $\mathcal{O}(r^{-1})$ while all the higher $\ell$ modes fall off as $\mathcal{O}(r^{-\ell-1})$. Analogously, all the $\bgkk_\ell{}^0$  modes fall off with the rate $\mathcal{O}(r^{-\ell-1})$, whereas the $\bgKK_\ell{}^0$ modes as $\mathcal{O}(r^{-\ell-2})$.

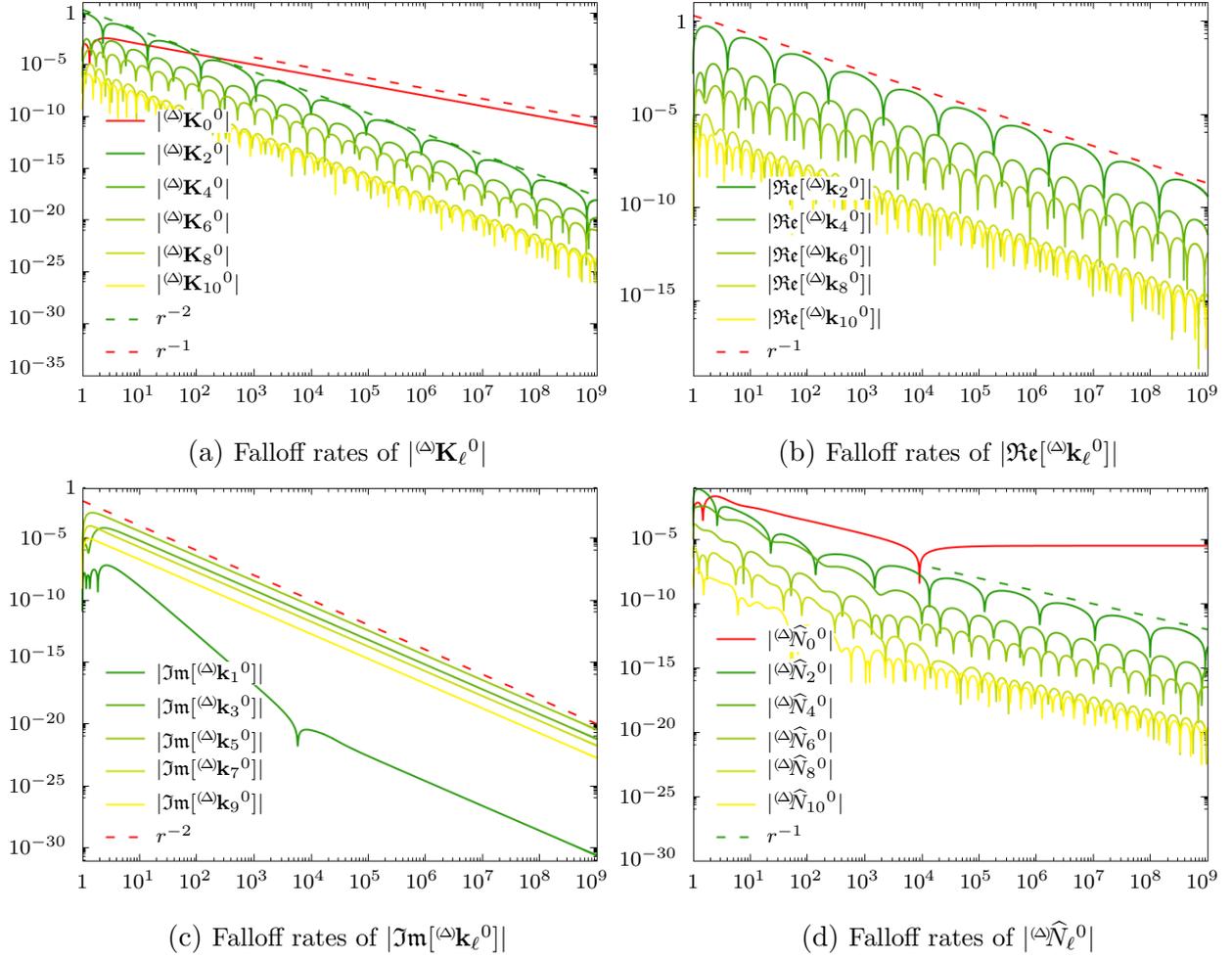
\begin{figure}[H]
		\begin{subfigure}[b]{.5\textwidth}
			\begin{overpic}[width=\textwidth]{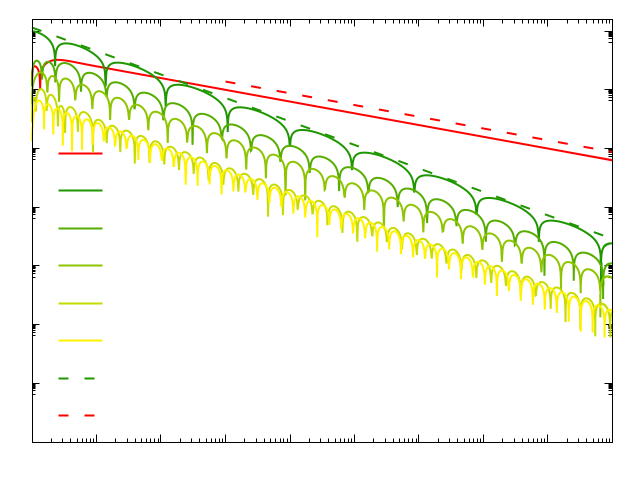}
				\put(4,1){\scriptsize$1$}
				\put(12,1){\scriptsize$10^1$}
				\put(22,1){\scriptsize$10^2$}
				\put(32,1){\scriptsize$10^3$}
				\put(42,1){\scriptsize$10^4$}
				\put(52,1){\scriptsize$10^5$}
				\put(62,1){\scriptsize$10^6$}
				\put(72,1){\scriptsize$10^7$}
				\put(82,1){\scriptsize$10^8$}
				\put(92,1){\scriptsize$10^9$}
				\put(2,69){\scriptsize$1$}
				\put(-6,60){\scriptsize$10^{-5}$}
				\put(-8,51){\scriptsize$10^{-10}$}
				\put(-8,42){\scriptsize$10^{-15}$}
				\put(-8,33){\scriptsize$10^{-20}$}
				\put(-8,24){\scriptsize$10^{-25}$}
				\put(-8,15){\scriptsize$10^{-30}$}
				\put(-8,6){\scriptsize$10^{-35}$}
				\put(18,50){\scriptsize\fboxsep0pt\colorbox{white}{$|\dKK_0{}^0|$}}
				\put(18,44){\scriptsize\fboxsep0pt\colorbox{white}{$|\dKK_2{}^0|$}}
				\put(18,38){\scriptsize$|\dKK_4{}^0|$}
				\put(18,32){\scriptsize$|\dKK_6{}^0|$}
				\put(18,26){\scriptsize$|\dKK_8{}^0|$}
				\put(18,21){\scriptsize$|\dKK_{10}{}^0|$}
				\put(18,15){\scriptsize$r^{-2}$}
				\put(18,9){\scriptsize$r^{-1}$}
			\end{overpic}
			\caption{\footnotesize Falloff rates of $|\dKK_\ell{}^0|$}
			\label{fig:dphKK}
		\end{subfigure}\hspace{.5cm}
		\begin{subfigure}[b]{.5\textwidth}
			\begin{overpic}[width=\textwidth]{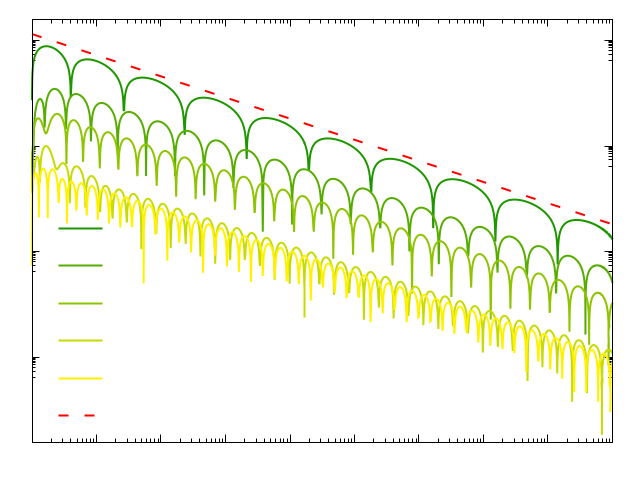}
				\put(4,1){\scriptsize$1$}
				\put(12,1){\scriptsize$10^1$}
				\put(22,1){\scriptsize$10^2$}
				\put(32,1){\scriptsize$10^3$}
				\put(42,1){\scriptsize$10^4$}
				\put(52,1){\scriptsize$10^5$}
				\put(62,1){\scriptsize$10^6$}
				\put(72,1){\scriptsize$10^7$}
				\put(82,1){\scriptsize$10^8$}
				\put(92,1){\scriptsize$10^9$}
				\put(2,67.5){\scriptsize$1$}
				\put(-6,51){\scriptsize$10^{-5}$}
				\put(-8,35){\scriptsize$10^{-10}$}
				\put(-8,19){\scriptsize$10^{-15}$}
				\put(18,38){\scriptsize\fboxsep0pt\colorbox{white}{$|\Re[\dkk_2{}^0]|$}}
				\put(18,32){\scriptsize\fboxsep0pt\colorbox{white}{$|\Re[\dkk_4{}^0]|$}}
				\put(18,26){\scriptsize$|\Re[\dkk_6{}^0]|$}
				\put(18,21){\scriptsize$|\Re[\dkk_{8}{}^0]|$}
				\put(18,15){\scriptsize$|\Re[\dkk_{10}{}^0]|$}
				\put(18,9){\scriptsize$r^{-1}$}
			\end{overpic}
			\caption{\footnotesize Falloff rates of $|\Re[\dkk_\ell{}^0]|$}
			\label{fig:dphRekk}
		\end{subfigure}\\
		\begin{subfigure}[b]{.5\textwidth}
			\begin{overpic}[width=\textwidth]{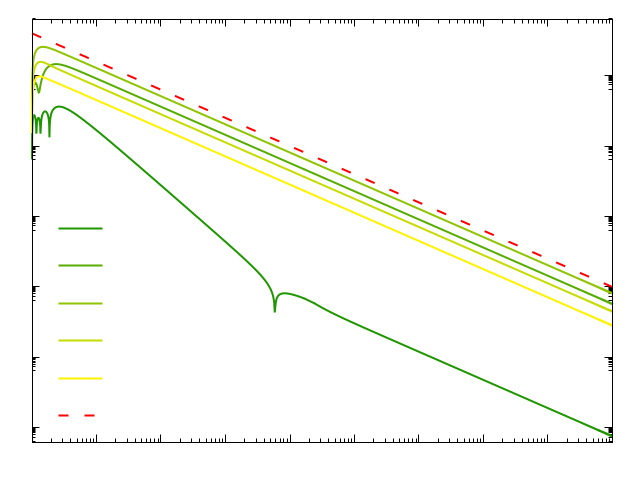}
				\put(4,1){\scriptsize$1$}
				\put(12,1){\scriptsize$10^1$}
				\put(22,1){\scriptsize$10^2$}
				\put(32,1){\scriptsize$10^3$}
				\put(42,1){\scriptsize$10^4$}
				\put(52,1){\scriptsize$10^5$}
				\put(62,1){\scriptsize$10^6$}
				\put(72,1){\scriptsize$10^7$}
				\put(82,1){\scriptsize$10^8$}
				\put(92,1){\scriptsize$10^9$}
				\put(2,71){\scriptsize$1$}
				\put(-6,62){\scriptsize$10^{-5}$}
				\put(-8,51){\scriptsize$10^{-10}$}
				\put(-8,41){\scriptsize$10^{-15}$}
				\put(-8,29){\scriptsize$10^{-20}$}
				\put(-8,18){\scriptsize$10^{-25}$}
				\put(-8,7){\scriptsize$10^{-30}$}
				\put(18,38){\scriptsize\fboxsep0pt\colorbox{white}{$|\Im[\dkk_1{}^0]|$}}
				\put(18,32){\scriptsize\fboxsep0pt\colorbox{white}{$|\Im[\dkk_3{}^0]|$}}
				\put(18,26){\scriptsize$|\Im[\dkk_5{}^0]|$}
				\put(18,21){\scriptsize$|\Im[\dkk_7{}^0]|$}
				\put(18,15){\scriptsize$|\Im[\dkk_9{}^0]|$}
				\put(18,9){\scriptsize$r^{-2}$}
			\end{overpic}
			\caption{\footnotesize Falloff rates of $|\Im[\dkk_\ell{}^0]|$}
			\label{fig:dphImkk}
		\end{subfigure}\hspace{.5cm}
		\begin{subfigure}[b]{.5\textwidth}
			\begin{overpic}[width=\textwidth]{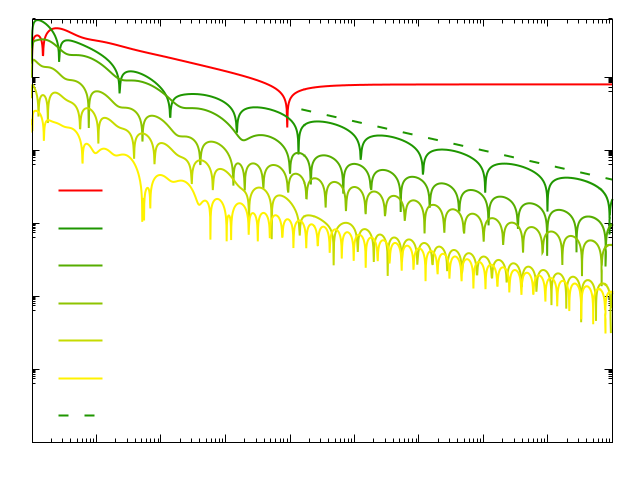}
				\put(4,1){\scriptsize$1$}
				\put(12,1){\scriptsize$10^1$}
				\put(22,1){\scriptsize$10^2$}
				\put(32,1){\scriptsize$10^3$}
				\put(42,1){\scriptsize$10^4$}
				\put(52,1){\scriptsize$10^5$}
				\put(62,1){\scriptsize$10^6$}
				\put(72,1){\scriptsize$10^7$}
				\put(82,1){\scriptsize$10^8$}
				\put(92,1){\scriptsize$10^9$}
				\put(-6,62){\scriptsize$10^{-5}$}
				\put(-8,51){\scriptsize$10^{-10}$}
				\put(-8,39){\scriptsize$10^{-15}$}
				\put(-8,28){\scriptsize$10^{-20}$}
				\put(-8,16){\scriptsize$10^{-25}$}
				\put(-8,5){\scriptsize$10^{-30}$}
				\put(18,44){\scriptsize\fboxsep0pt\colorbox{white}{$|\dNh_0{}^0|$}}
				\put(18,38){\scriptsize\fboxsep0pt\colorbox{white}{$|\dNh_2{}^0|$}}
				\put(18,32){\scriptsize$|\dNh_4{}^0|$}
				\put(18,26){\scriptsize$|\dNh_6{}^0|$}
				\put(18,21){\scriptsize$|\dNh_8{}^0|$}
				\put(18,15){\scriptsize$|\dNh_{10}{}^0|$}
				\put(18,9){\scriptsize$r^{-1}$}
			\end{overpic}
			\caption{\footnotesize Falloff rates of $|\dNh_\ell{}^0|$}
			\label{fig:dphNh}
		\end{subfigure}
		\caption{\footnotesize The falloff rates of the only nontrivial modes of the constrained variables obtained by integrating the parabolic-hyperbolic system for strictly near-Kerr initial data are depicted. It is transparent that the falloff rate of $\dKK_0{}^0$ or that of $\dNh_0{}^0$, individually,  are capable of excluding the conditions of even the weak asymptotic flatness to hold.}
		\label{fig:dphorig}
	\end{figure}

	In proceeding, note first that on all of the figures included in this paper, to be able to use log-log scales and thereby demonstrate the claimed falloff rates, we plot the absolute values of the nonlinear perturbations, along with some auxiliary lines helping the comparison with the expected rates. Specifically, in Fig.\,\ref{fig:dphorig}, various modes of the constrained fields, obtained by integrating the parabolic-hyperbolic equations, in the strictly near-Kerr case are plotted. While the monopole part $\dKK_0{}^0$ of $\dKK$ decays as $\mathcal{O}(r^{-1})$, all the higher $\ell$-modes fall off as $\mathcal{O}(r^{-2})$. The modes comprising the real part of $\dkk$ fall off as  $\mathcal{O}(r^{-1})$, whereas the modes in the imaginary part of $\dkk$ decay with the rate $\mathcal{O}(r^{-2})$. In Fig.\,\ref{fig:dphNh} it is transparent that  instead of its proper limit value $0$, the monopole part $\dNh_0{}^0$ tends to $3.4114\cdot10^{-6}$. By contrast, all the other modes of $\dNh$ fall off as  $\mathcal{O}(r^{-1})$. These observations underline that likewise, it occurred in the strictly near-Schwarzschild case, the asymptotic behavior of either $\dKK_0{}^0$ or $\dNh_0{}^0$, respectively, prevents even the weak asymptotic flatness of the corresponding strictly near-Kerr initial dataset.

\begin{figure}[H]
      \begin{subfigure}[b]{.5\textwidth}
        \begin{overpic}[width=\textwidth]{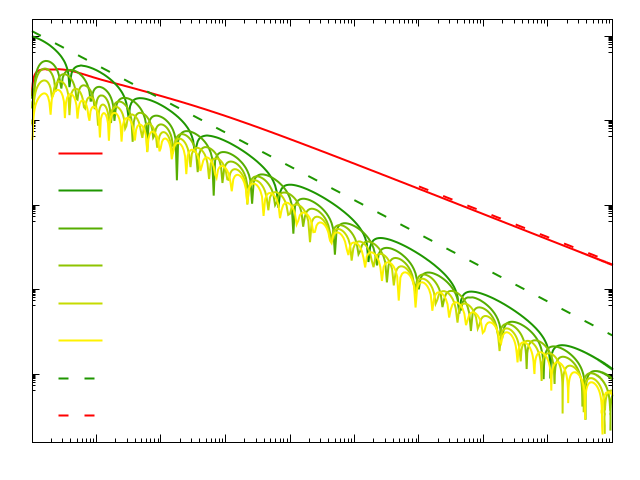}
          \put(4,1){\scriptsize$1$}
          \put(12,1){\scriptsize$10^1$}
          \put(22,1){\scriptsize$10^2$}
          \put(32,1){\scriptsize$10^3$}
          \put(42,1){\scriptsize$10^4$}
          \put(52,1){\scriptsize$10^5$}
          \put(62,1){\scriptsize$10^6$}
          \put(72,1){\scriptsize$10^7$}
          \put(82,1){\scriptsize$10^8$}
          \put(92,1){\scriptsize$10^9$}
          \put(2,68){\scriptsize$1$}
          \put(-6,55){\scriptsize$10^{-5}$}
          \put(-8,42){\scriptsize$10^{-10}$}
          \put(-8,29){\scriptsize$10^{-15}$}
          \put(-8,16){\scriptsize$10^{-20}$}
          \put(18,50){\scriptsize\fboxsep0pt\colorbox{white}{$|\dKK_0{}^0|$}}
          \put(18,44){\scriptsize\fboxsep0pt\colorbox{white}{$|\dKK_2{}^0|$}}
          \put(18,38){\scriptsize$|\dKK_4{}^0|$}
          \put(18,32){\scriptsize$|\dKK_6{}^0|$}
          \put(18,26){\scriptsize$|\dKK_8{}^0|$}
          \put(18,21){\scriptsize$|\dKK_{10}{}^0|$}
          \put(18,15){\scriptsize$r^{-2}$}
          \put(18,9){\scriptsize$r^{-3/2}$}
        \end{overpic}
        \caption{\footnotesize Falloff rates of $|\dKK_\ell{}^0|$}
        \label{fig:dahKK}
      \end{subfigure}\hspace{.5cm}
      \begin{subfigure}[b]{.5\textwidth}
        \begin{overpic}[width=\textwidth]{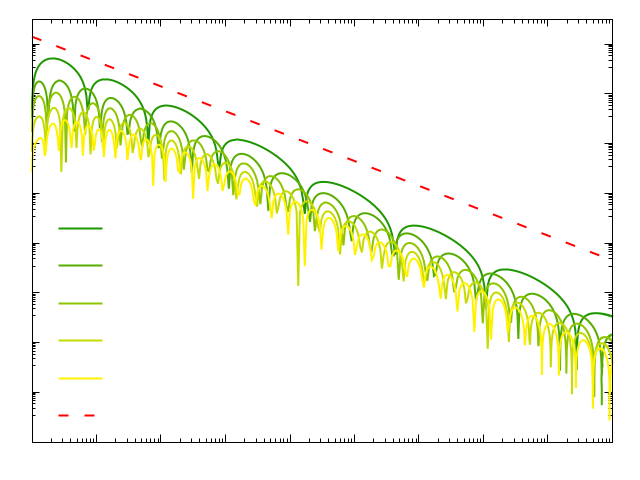}
          \put(4,1){\scriptsize$1$}
          \put(12,1){\scriptsize$10^1$}
          \put(22,1){\scriptsize$10^2$}
          \put(32,1){\scriptsize$10^3$}
          \put(42,1){\scriptsize$10^4$}
          \put(52,1){\scriptsize$10^5$}
          \put(62,1){\scriptsize$10^6$}
          \put(72,1){\scriptsize$10^7$}
          \put(82,1){\scriptsize$10^8$}
          \put(92,1){\scriptsize$10^9$}
          \put(2,67){\scriptsize$1$}
          \put(-6,59){\scriptsize$10^{-2}$}
          \put(-6,51){\scriptsize$10^{-4}$}
          \put(-6,43){\scriptsize$10^{-6}$}
          \put(-6,35){\scriptsize$10^{-8}$}
          \put(-8,28){\scriptsize$10^{-10}$}
          \put(-8,20){\scriptsize$10^{-12}$}
          \put(-8,12){\scriptsize$10^{-14}$}
          \put(-8,5){\scriptsize$10^{-16}$}
          \put(18,38){\scriptsize$|\Re[\dkk_2{}^0]|$}
          \put(18,32){\scriptsize$|\Re[\dkk_4{}^0]|$}
          \put(18,26){\scriptsize$|\Re[\dkk_6{}^0]|$}
          \put(18,21){\scriptsize$|\Re[\dkk_{8}{}^0]|$}
          \put(18,15){\scriptsize$|\Re[\dkk_{10}{}^0]|$}
          \put(18,9){\scriptsize$r^{-1}$}
        \end{overpic}
        \caption{\footnotesize Falloff rates of $|\Re[\dkk_\ell{}^0]|$}
        \label{fig:dahRekk}
      \end{subfigure}
  		\vskip0.2cm
      \begin{subfigure}[b]{.5\textwidth}
        \begin{overpic}[width=\textwidth]{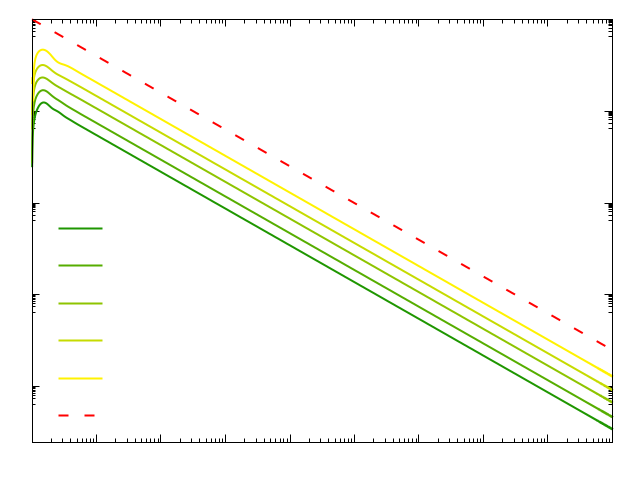}
          \put(4,1){\scriptsize$1$}
          \put(12,1){\scriptsize$10^1$}
          \put(22,1){\scriptsize$10^2$}
          \put(32,1){\scriptsize$10^3$}
          \put(42,1){\scriptsize$10^4$}
          \put(52,1){\scriptsize$10^5$}
          \put(62,1){\scriptsize$10^6$}
          \put(72,1){\scriptsize$10^7$}
          \put(82,1){\scriptsize$10^8$}
          \put(92,1){\scriptsize$10^9$}
          \put(-6,70){\scriptsize$10^{-5}$}
          \put(-8,56){\scriptsize$10^{-10}$}
          \put(-8,42){\scriptsize$10^{-15}$}
          \put(-8,28){\scriptsize$10^{-20}$}
          \put(-8,13){\scriptsize$10^{-25}$}
          \put(18,38){\scriptsize\fboxsep0pt\colorbox{white}{$|\Im[\dkk_1{}^0]|$}}
          \put(18,32){\scriptsize$|\Im[\dkk_3{}^0]|$}
          \put(18,26){\scriptsize$|\Im[\dkk_5{}^0]|$}
          \put(18,21){\scriptsize$|\Im[\dkk_7{}^0]|$}
          \put(18,15){\scriptsize$|\Im[\dkk_9{}^0]|$}
          \put(18,9){\scriptsize$r^{-2}$}
        \end{overpic}
        \caption{\footnotesize Falloff rates of $|\Im[\dkk_\ell{}^0]|$}
        \label{fig:dahImkk}
      \end{subfigure}\hspace{.5cm}
      \begin{subfigure}[b]{.5\textwidth}
      	\begin{overpic}[width=\textwidth]{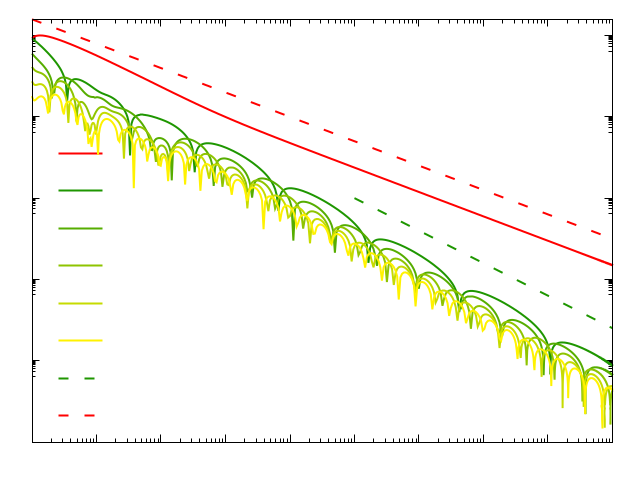}
      		\put(4,1){\scriptsize$1$}
      		\put(12,1){\scriptsize$10^1$}
      		\put(22,1){\scriptsize$10^2$}
      		\put(32,1){\scriptsize$10^3$}
      		\put(42,1){\scriptsize$10^4$}
      		\put(52,1){\scriptsize$10^5$}
      		\put(62,1){\scriptsize$10^6$}
      		\put(72,1){\scriptsize$10^7$}
      		\put(82,1){\scriptsize$10^8$}
      		\put(92,1){\scriptsize$10^9$}
          \put(2,68.5){\scriptsize$1$}
      		\put(-6,56){\scriptsize$10^{-5}$}
      		\put(-8,43){\scriptsize$10^{-10}$}
      		\put(-8,30){\scriptsize$10^{-15}$}
      		\put(-8,17.5){\scriptsize$10^{-20}$}
      		\put(-8,5){\scriptsize$10^{-25}$}
          \put(18,50){\scriptsize\fboxsep0pt\colorbox{white}{$|\kkappa_0{}^0|$}}
          \put(18,44){\scriptsize\fboxsep0pt\colorbox{white}{$|\kkappa_2{}^0|$}}
          \put(18,38){\scriptsize$|\kkappa_4{}^0|$}
          \put(18,32){\scriptsize$|\kkappa_6{}^0|$}
          \put(18,26){\scriptsize$|\kkappa_8{}^0|$}
          \put(18,21){\scriptsize$|\kkappa_{10}{}^0|$}
          \put(18,15){\scriptsize$r^{-2}$}
          \put(18,9){\scriptsize$r^{-3/2}$}
      	\end{overpic}
      	\caption{\footnotesize Falloff rates of $|\kkappa_\ell{}^0|$}
      	\label{fig:dahImkk}
      \end{subfigure}
  		\vskip-0.2cm
      \caption{\footnotesize The falloff rates of the only nontrivial modes of the constrained variables yielded by the algebraic-hyperbolic equations for strictly near-Kerr initial data are shown. It is visible that apart from the monopole part $\dKK_0{}^0$ of $\dKK$ and $\kkappa_0{}^0$ of $\kkappa$, each of the other modes falls off at a rate that is even faster than required by conditions of strong asymptotic flatness.}
      \label{fig:dahorig}
    \end{figure}
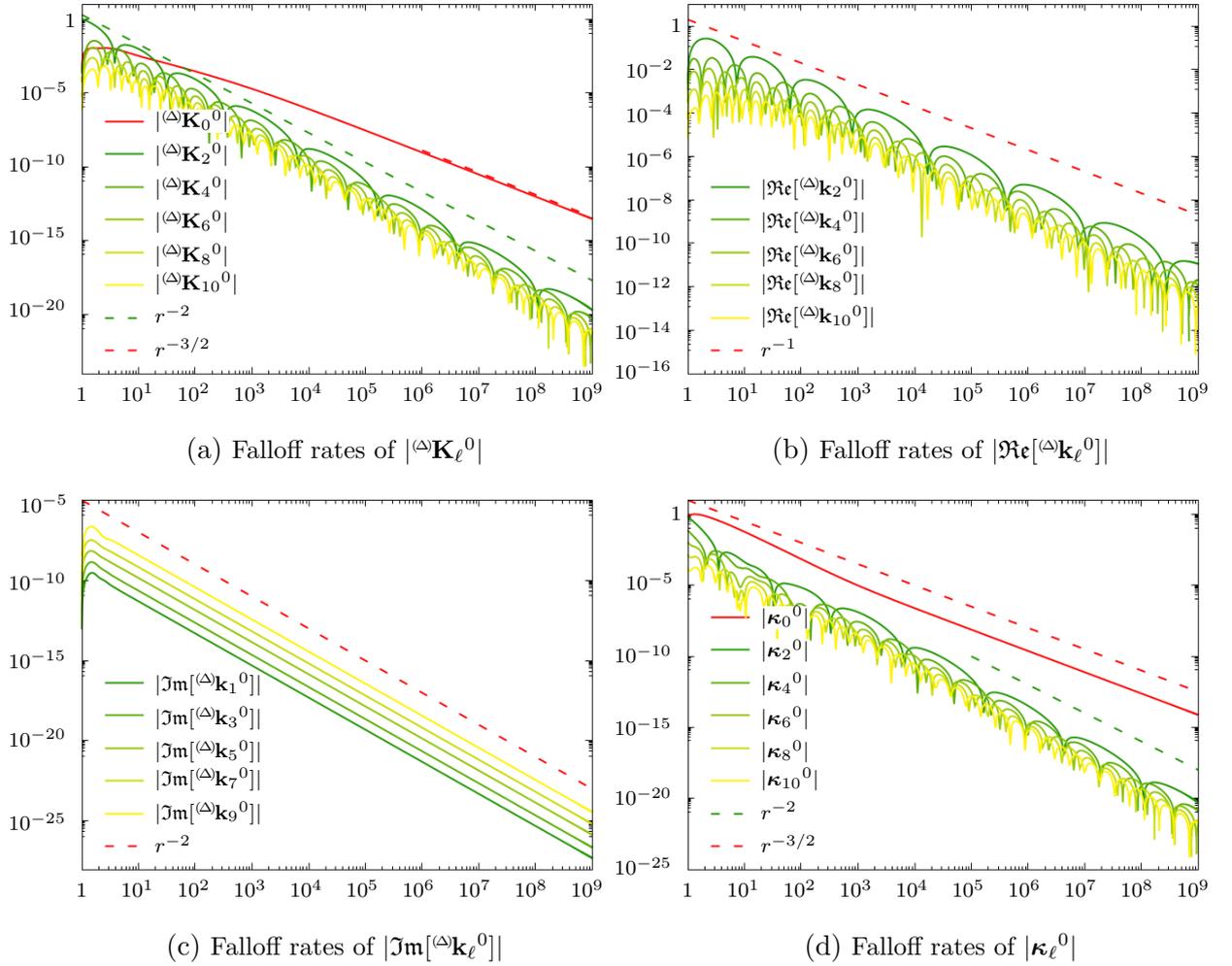

	Analogously, in Fig.\,\ref{fig:dahorig}, various modes of the constrained variables, obtained by integrating the algebraic-hyperbolic system, in the strictly near-Kerr case are shown. It is transparent that $\dKK_0{}^0$ decays with the rate $\mathcal{O}(r^{-3/2})$. Notably, all the $\KK_\ell{}^0$ higher $\ell$ modes fall off somewhat faster than $\mathcal{O}(r^{-2})$, with a rate close to $\mathcal{O}(r^{-2.2})$. Similarly, all the modes comprising the real part of $\dkk$ also decay faster than $\mathcal{O}(r^{-1})$, with a rate close to $\mathcal{O}(r^{-1.2})$, whereas the modes in the imaginary part of $\dkk$ fall off with the rate $\mathcal{O}(r^{-2})$. As a consequence of the slow decay rate of $\dKK$, the solution to the algebraic Hamiltonian constraint, $\kkappa$, also has the slow $\mathcal{O}(r^{-3/2})$ decay rate for the monopole, $\kkappa_0{}^0$, with $\ell>0$ modes decaying slightly faster than $\mathcal{O}(r^{-2})$. Accordingly, likewise in the strictly near-Schwarzschild case, the only modes violating both strong and weak asymptotic flatness are $\dKK_0{}^0$ and $\kkappa_0{}^0$.

	Our investigation reported in the panels of Figs.\,\ref{fig:dphorig} and \ref{fig:dahorig} confirm that, in the case of the strictly near-Kerr initial datasets, apart from the monopole parts of $\dKK$ and $\dNh$, in case of the parabolic-hyperbolic equations, and apart from the monopole parts of $\dKK$ and $\kkappa$ in case of the algebraic-hyperbolic equations, respectively, all the other modes decay with a rate that would allow even strong asymptotic flatness.

  \subsection{Applying the new method to the parabolic-hyperbolic system}
  \label{sec:phm}

	As was mentioned in the Introduction, Beyer {\it et al.}\,\cite{Beyer:2020zlo} invented a method that applies to the constraints' parabolic-hyperbolic formulation, allowing them to produce strongly asymptotically flat near-Schwarzschild initial datasets. This was done by setting the freely specifiable variable $\kkappa$ to be proportional to $\KK$. Though using the relation $\kkappa = \mathcal{R}\,\KK$ allows one to produce strongly asymptotically flat near-Schwarzschild initial datasets, it also affects the principal part of the parabolic-hyperbolic system as it also contains tangential derivatives of $\kkappa$. The inequality $\mathcal{R}>-\tfrac12$ had to be imposed on the proportionality factor to ensure well posedness of the system proposed in \cite{Beyer:2020zlo}.

	It was also mentioned earlier that one of our aims is to introduce a method that can be applied on equal footing to the parabolic-hyperbolic and the algebraic-hyperbolic systems such that no change in the principal parts occurs. This idea was motivated by our observations in \cite{Csukas:2019qco} concerning strictly near-Schwarzschild initial datasets and by the results covered in the previous section relevant to strictly near-Kerr initial datasets. Both of these mode-by-mode investigations pointed to the fact that they are indeed the monopole parts of $\dKK$ and $\dNh$ in the case of the parabolic-hyperbolic equations, and it is the monopole part of $\dKK$ in the case of the algebraic-hyperbolic equations that are needed to be adjusted. This immediately raises the question if it is possible to get the desired falloff rates by restricting only the monopole parts of some of the freely specifiable variables such that for both the parabolic-hyperbolic and the algebraic-hyperbolic systems, we can get asymptotically flat near-Kerr initial data configurations.

	It turned out that essentially by demanding the simple variant of the choice of Beyer {\it et al.}\,\cite{Beyer:2020zlo} given below helps not only in getting strongly asymptotically flat solutions to both of the alternative evolutionary forms of the constraints but also we can have a more flexible control on the falloff rates, and, in particular, we can produce weakly asymptotically flat initial data with any desired decay rate.

	In the parabolic hyperbolic case we restrict only the monopole part of the freely specifiable field $\kkappa$ by setting
	\begin{equation}
		\label{eq:newscR}
		\kkappa_0{}^0=\alpha\left[\mathscr{R}\,\KK\right]_0{}^0\,,\quad {\rm with} \quad \mathscr{R} = \bgkkappa/\bgKK\,,
	\end{equation}
	where $\alpha$ is a positive real number. As we shall see below, the role of the parameter  $\alpha$ is to fine-tune the falloff rate of the monopole part of $\KK$. Note also that the choice $\mathscr{R} = \bgkkappa/\bgKK$ guarantees recovering the Kerr-Schild data on $\Sigma$ with choosing $\alpha=1$ and trivial initial data for the nonlinear perturbations of the constrained variables.

	In the parabolic-hyperbolic case, we also assume that, apart from the monopole part of $\kkappa$, all the other modes remain intact, i.e.,
	\begin{equation}
		\kkappa_\ell{}^0\big|_{\ell>0}=\bgkkappa_\ell{}^0\,.
	\end{equation}

	One of our primary motivations for controlling only the monopole part of the freely specifiable variable $\kkappa$ by using \eqref{eq:newscR} was that for the tangential derivatives of $\kkappa$ the relation $\eth\kkappa=\eth\bgkkappa$ holds. Accordingly, the principal part of the parabolic-hyperbolic equations, \eqref{eq:phN}, \eqref{eq:phk}, and \eqref{eq:phK}, remains intact, so it is reasonable to argue, that the local well posedness of the system follows from results covered in \cite{Racz:2015mfa}.

\subsubsection{Outline in the spherically symmetric setup}

	In spherical symmetry,\footnote{As in the spherically symmetric case, the variables have only monopole parts; the $\ell,m$ indices are suppressed.} using the Schwarzschild background, the parabolic-hyperbolic equations \eqref{eq:phN} and \eqref{eq:phK} simplify to the system
	\begin{align}
		\tfrac{2}{r}\tfrac{\mathrm{d}\Nh}{\mathrm{d}r}&=\tfrac{1}{r^2}\,\Nh-\tfrac{1}{2}\,\Nh^3\Big(\tfrac{2}{r^2}+2\,\kkappa\,\KK+\tfrac{1}{2}\,\KK^2\Big)\,,\label{eq:phmNsph}\\
		\tfrac{\mathrm{d}\KK}{\mathrm{d}r}&=\tfrac{2}{r}\,\Big(\kkappa-\tfrac{1}{2}\KK\Big)\,.
		\label{eq:phmKsph}
	\end{align}
	Substituting \eqref{eq:newscR} with  $\mathscr{R}=-(r+M)/(2\,(r+2\,M))$ (relevant for the  Schwarzschild background) immediately gives the solution to \eqref{eq:phmKsph} as
	\begin{equation}
	 \label{eq:phmKKsol}
	 \KK=\frac{C_1}{r^{1+\alpha}(1+2M/r)^{\alpha/2}}\,,
	\end{equation}
	possessing the asymptotic expansion
	\begin{equation}
	 \label{eq:phmKKasym}
	 \KK\sim \frac{C_1}{r^{1+\alpha}}-\frac{C_1 M \alpha}{r^{2+\alpha}}+\mathcal{O}\left(r^{-(3+\alpha)}\right)\,,
	\end{equation}
  where $C_1$ is a constant of integration.
 	By virtue of \eqref{eq:phmKKasym} the parameter $\alpha$, applied in \eqref{eq:newscR}, allows a straightforward fine-tuning of the asymptotic falloff rate of $\KK$. To guarantee $\KK$ to fit the requirements in weak asymptotic flatness, $\alpha>1/2$ should hold. If $\alpha=1$, the falloff rate in \eqref{eq:phmKKasym} is compatible with the requirements in strong asymptotic flatness. In the spherically symmetric case, we could also choose faster falloff rates by setting $\alpha>1$. Note, however, that such a rapid decay does not occur in the nonspherically symmetric case because there are intimate couplings of $\dKK_0{}^0$ and many of the background fields with falloff rate $\mathcal{O}(r^{-2})$. Thereby, in the generic case, one should not expect faster than  $\mathcal{O}(r^{-2})$  falloff rate for  $\dKK_0{}^0$.

  Substituting \eqref{eq:phmKKsol} into \eqref{eq:phmNsph} we get
  \begin{equation}
   \Nh=\left[1+\tfrac{C_2}{r}+\tfrac{C_1{}^2}{4}r^{-2\alpha}\left(1+\tfrac{2M}{r}\right)\right]^{-1/2}
  \end{equation}
  with the desired asymptotic value $1$, where $C_2$ is another constant of integration. It is transparent that if $\alpha$ was smaller than $1/2$, then the term $-\tfrac{C_1{}^2}{8r^{2\alpha}}$ could dominate, and whence $\Nh$ would fail to fall off with the rate $\mathcal{O}(r^{-1})$ that is required by the weak and strong forms of asymptotic flatness in \eqref{eq:weak-asymp} and \eqref{eq:stro-asymp}, respectively. While the choice of the borderline value $\alpha=1/2$ would fit $\Nh$, it is only the interval $\alpha>1/2$ that ensures the decay rates are, simultaneously for $\KK$ and $\Nh$, compatible with the weak and strong forms of asymptotic flatness in \eqref{eq:weak-asymp} and \eqref{eq:stro-asymp}, respectively.

\subsubsection{Strongly asymptotically flat numerical solution}

\begin{figure}[H]
    \begin{subfigure}[b]{.5\textwidth}
      \begin{overpic}[width=\textwidth]{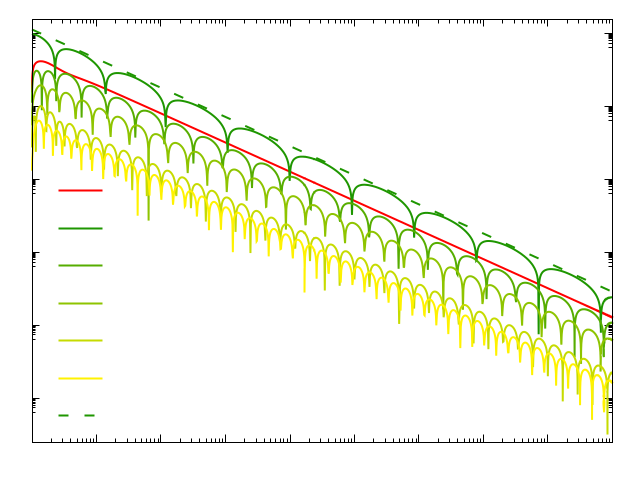}
        \put(4,1){\scriptsize$1$}
        \put(12,1){\scriptsize$10^1$}
        \put(22,1){\scriptsize$10^2$}
        \put(32,1){\scriptsize$10^3$}
        \put(42,1){\scriptsize$10^4$}
        \put(52,1){\scriptsize$10^5$}
        \put(62,1){\scriptsize$10^6$}
        \put(72,1){\scriptsize$10^7$}
        \put(82,1){\scriptsize$10^8$}
        \put(92,1){\scriptsize$10^9$}
        \put(2,68){\scriptsize$1$}
        \put(-6,57){\scriptsize$10^{-5}$}
        \put(-8,45){\scriptsize$10^{-10}$}
        \put(-8,34){\scriptsize$10^{-15}$}
        \put(-8,23){\scriptsize$10^{-20}$}
        \put(-8,12){\scriptsize$10^{-25}$}
        \put(18,44){\scriptsize\fboxsep0pt\colorbox{white}{$|\dKK_0{}^0|$}}
        \put(18,38){\scriptsize\fboxsep0pt\colorbox{white}{$|\dKK_2{}^0|$}}
        \put(18,32){\scriptsize$|\dKK_4{}^0|$}
        \put(18,26){\scriptsize$|\dKK_6{}^0|$}
        \put(18,21){\scriptsize$|\dKK_8{}^0|$}
        \put(18,15){\scriptsize$|\dKK_{10}{}^0|$}
        \put(18,9){\scriptsize$r^{-2}$}
      \end{overpic}
      \caption{\footnotesize Falloff rates of $|\dKK_\ell{}^0|$}
      \label{fig:phbrm2KK}
    \end{subfigure}\hspace{.5cm}
    \begin{subfigure}[b]{.5\textwidth}
      \begin{overpic}[width=\textwidth]{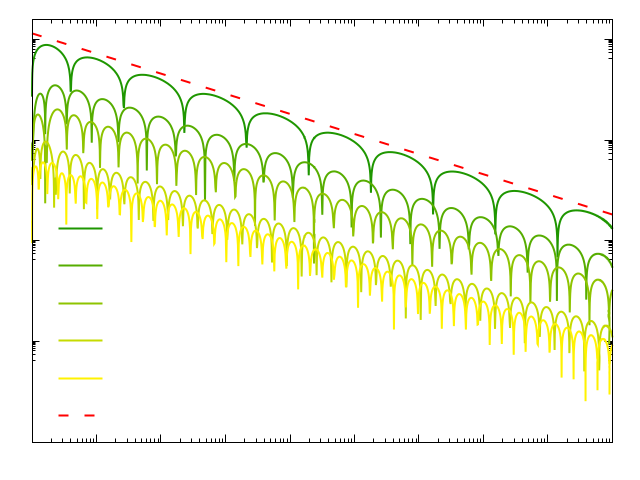}
        \put(4,1){\scriptsize$1$}
        \put(12,1){\scriptsize$10^1$}
        \put(22,1){\scriptsize$10^2$}
        \put(32,1){\scriptsize$10^3$}
        \put(42,1){\scriptsize$10^4$}
        \put(52,1){\scriptsize$10^5$}
        \put(62,1){\scriptsize$10^6$}
        \put(72,1){\scriptsize$10^7$}
        \put(82,1){\scriptsize$10^8$}
        \put(92,1){\scriptsize$10^9$}
        \put(2,67.5){\scriptsize$1$}
        \put(-6,52){\scriptsize$10^{-5}$}
        \put(-8,36){\scriptsize$10^{-10}$}
        \put(-8,21){\scriptsize$10^{-15}$}
        \put(18,38){\scriptsize\fboxsep0pt\colorbox{white}{$|\Re[\dkk_2{}^0]|$}}
        \put(18,32){\scriptsize\fboxsep0pt\colorbox{white}{$|\Re[\dkk_4{}^0]|$}}
        \put(18,26){\scriptsize$|\Re[\dkk_6{}^0]|$}
        \put(18,21){\scriptsize$|\Re[\dkk_{8}{}^0]|$}
        \put(18,15){\scriptsize$|\Re[\dkk_{10}{}^0]|$}
        \put(18,9){\scriptsize$r^{-1}$}
      \end{overpic}
      \caption{\footnotesize Falloff rates of $|\Re[\dkk_\ell{}^0]|$}
      \label{fig:phbrm2Rekk}
    \end{subfigure}
		\vskip0.2cm
    \begin{subfigure}[b]{.5\textwidth}
      \begin{overpic}[width=\textwidth]{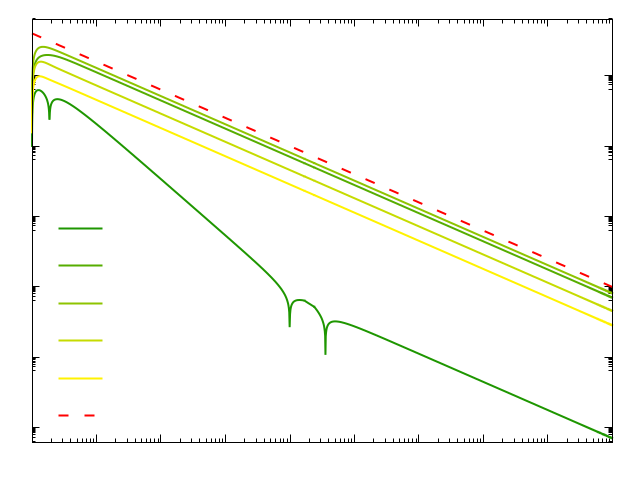}
        \put(4,1){\scriptsize$1$}
        \put(12,1){\scriptsize$10^1$}
        \put(22,1){\scriptsize$10^2$}
        \put(32,1){\scriptsize$10^3$}
        \put(42,1){\scriptsize$10^4$}
        \put(52,1){\scriptsize$10^5$}
        \put(62,1){\scriptsize$10^6$}
        \put(72,1){\scriptsize$10^7$}
        \put(82,1){\scriptsize$10^8$}
        \put(92,1){\scriptsize$10^9$}
        \put(2,71){\scriptsize$1$}
        \put(-6,62){\scriptsize$10^{-5}$}
        \put(-8,51){\scriptsize$10^{-10}$}
        \put(-8,41){\scriptsize$10^{-15}$}
        \put(-8,29){\scriptsize$10^{-20}$}
        \put(-8,18){\scriptsize$10^{-25}$}
        \put(-8,7){\scriptsize$10^{-30}$}
        \put(18,38){\scriptsize\fboxsep0pt\colorbox{white}{$|\Im[\dkk_1{}^0]|$}}
        \put(18,32){\scriptsize\fboxsep0pt\colorbox{white}{$|\Im[\dkk_3{}^0]|$}}
        \put(18,26){\scriptsize$|\Im[\dkk_5{}^0]|$}
        \put(18,21){\scriptsize$|\Im[\dkk_7{}^0]|$}
        \put(18,15){\scriptsize$|\Im[\dkk_9{}^0]|$}
        \put(18,9){\scriptsize$r^{-2}$}
      \end{overpic}
      \caption{\footnotesize Falloff rates of $|\Im[\dkk_\ell{}^0]|$}
      \label{fig:phbrm2Imkk}
    \end{subfigure}\hspace{.5cm}
    \begin{subfigure}[b]{.5\textwidth}
      \begin{overpic}[width=\textwidth]{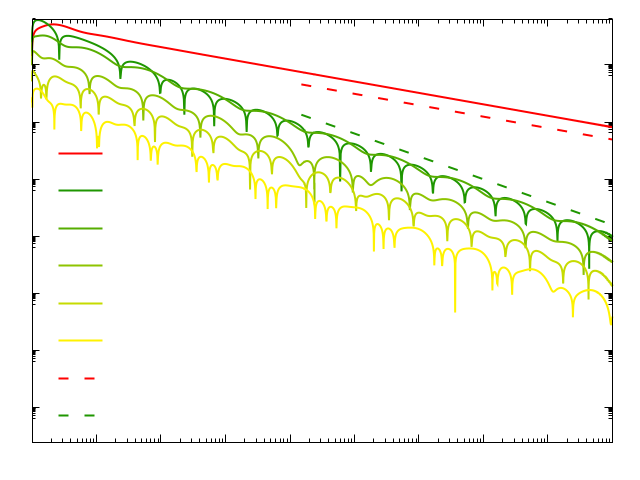}
        \put(4,1){\scriptsize$1$}
        \put(12,1){\scriptsize$10^1$}
        \put(22,1){\scriptsize$10^2$}
        \put(32,1){\scriptsize$10^3$}
        \put(42,1){\scriptsize$10^4$}
        \put(52,1){\scriptsize$10^5$}
        \put(62,1){\scriptsize$10^6$}
        \put(72,1){\scriptsize$10^7$}
        \put(82,1){\scriptsize$10^8$}
        \put(92,1){\scriptsize$10^9$}
        \put(-6,64){\scriptsize$10^{-5}$}
        \put(-8,55){\scriptsize$10^{-10}$}
        \put(-8,46){\scriptsize$10^{-15}$}
        \put(-8,37){\scriptsize$10^{-20}$}
        \put(-8,28){\scriptsize$10^{-25}$}
        \put(-8,19){\scriptsize$10^{-30}$}
        \put(-8,10){\scriptsize$10^{-35}$}
        \put(18,50){\scriptsize\fboxsep0pt\colorbox{white}{$|\dNh_0{}^0|$}}
        \put(18,44){\scriptsize\fboxsep0pt\colorbox{white}{$|\dNh_2{}^0|$}}
        \put(18,38){\scriptsize\fboxsep0pt\colorbox{white}{$|\dNh_4{}^0|$}}
        \put(18,32){\scriptsize$|\dNh_6{}^0|$}
        \put(18,26){\scriptsize$|\dNh_8{}^0|$}
        \put(18,21){\scriptsize$|\dNh_{10}{}^0|$}
        \put(18,15){\scriptsize$r^{-1}$}
        \put(18,9){\scriptsize$r^{-2}$}
      \end{overpic}
      \caption{\footnotesize Falloff rates of $|\dNh_\ell{}^0|$}
      \label{fig:phbrm2Nh}
    \end{subfigure}
    \caption{\footnotesize The falloff rates of the only nontrivial modes of the constrained variables yielded by the parabolic-hyperbolic equations for near-Kerr initial data, and applying \eqref{eq:newscR} with $\alpha=1$, are plotted. Note that each mode falls off in accordance with the conditions in strong asymptotic flatness.}
    \label{fig:phbrm2}
  \end{figure}
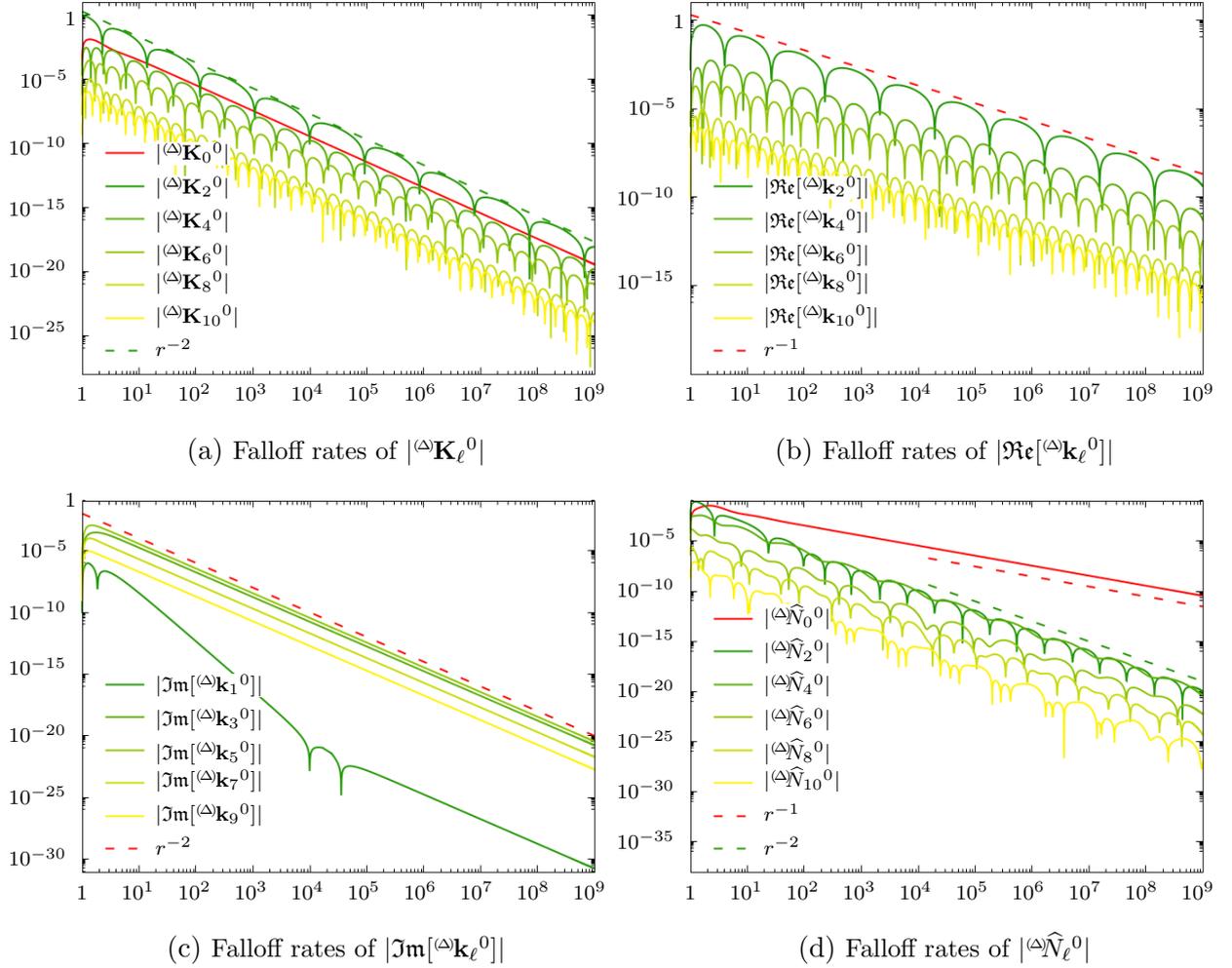

  In Fig.\,\ref{fig:phbrm2} the falloff rates of the constraint fields $\dKK, \dkk$, and $\dNh$, yielded by integrating the parabolic-hyperbolic equations \eqref{eq:phN}, \eqref{eq:phk}, and \eqref{eq:phK} for near-Kerr initial data, and applying \eqref{eq:newscR} with $\alpha=1$, are depicted. In this case each of the $\dKK_\ell{}^0$ modes falls off as $\mathcal{O}(r^{-2})$, $\Re[\dkk_\ell{}^0]$ as $\mathcal{O}(r^{-1})$ with $\Im[\dkk_\ell{}^0]$ going as $\mathcal{O}(r^{-2})$. Finally, as it is desired, $\dNh_0{}^0$ tends to $0$ with the rate $\mathcal{O}(r^{-1})$, whereas all the higher $\ell$ modes decay as $\mathcal{O}(r^{-2})$. Note that all of these falloff rates, apart from that of $\dkk$, are the same as reported in \cite{Csukas:2019qco}, relevant for near-Schwarzschild initial data using the method proposed in \cite{Beyer:2020zlo}. In that case $\dkk$ was purely real and each mode decayed at the rate $\mathcal{O}(r^{-2})$. In contrast, in the present case only the imaginary part retains this fast decay rate, whereas the decay of the real part is slower in  consequence of using \eqref{eq:newscR} instead of the proposal in \cite{Beyer:2020zlo}. Note however that each of the falloff rates indicated in Fig.\,\ref{fig:phbrm2} are compatible with the conditions of strong asymptotic flatness in \eqref{eq:stro-asymp} as we desired to show.

\subsubsection{Weakly asymptotically flat numerical solution}

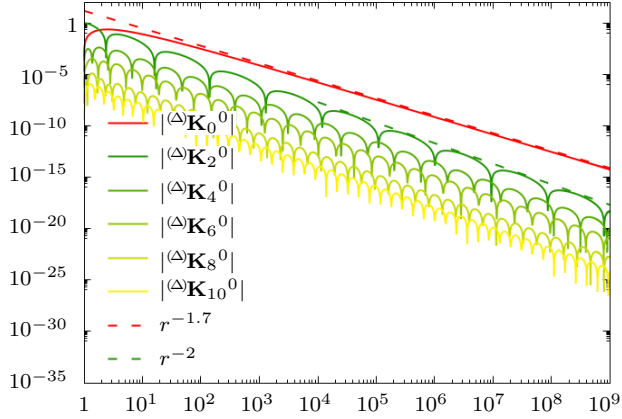
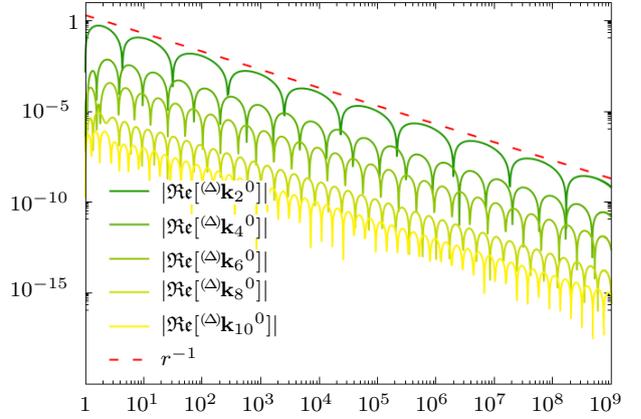
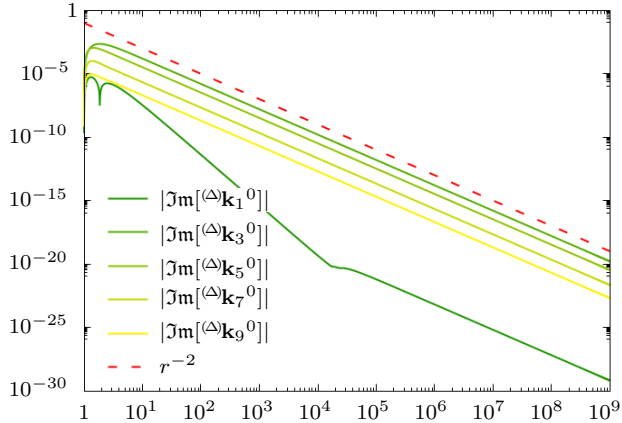
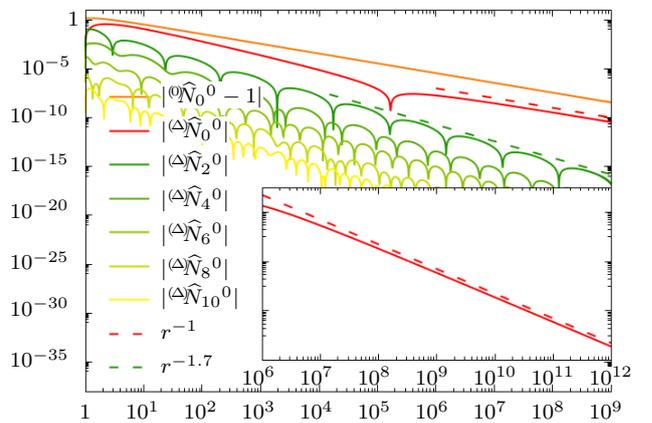
\begin{figure}[H]
    \begin{subfigure}[b]{.5\textwidth}
      \begin{overpic}[width=\textwidth]{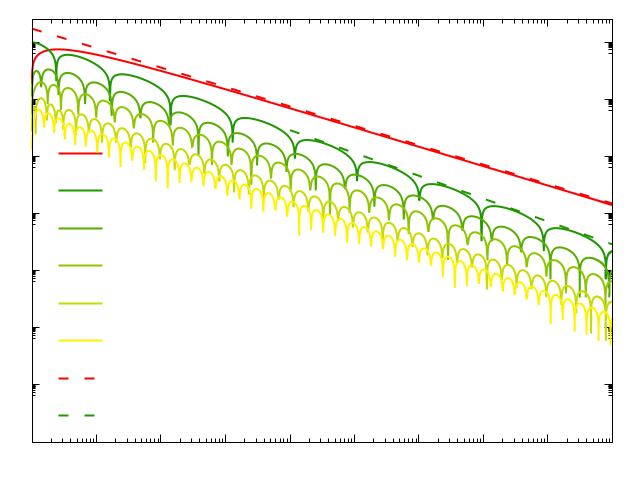}
        \put(4,1){\scriptsize$1$}
        \put(12,1){\scriptsize$10^1$}
        \put(22,1){\scriptsize$10^2$}
        \put(32,1){\scriptsize$10^3$}
        \put(42,1){\scriptsize$10^4$}
        \put(52,1){\scriptsize$10^5$}
        \put(62,1){\scriptsize$10^6$}
        \put(72,1){\scriptsize$10^7$}
        \put(82,1){\scriptsize$10^8$}
        \put(92,1){\scriptsize$10^9$}
        \put(2,67){\scriptsize$1$}
        \put(-6,58){\scriptsize$10^{-5}$}
        \put(-8,49){\scriptsize$10^{-10}$}
        \put(-8,41){\scriptsize$10^{-15}$}
        \put(-8,32){\scriptsize$10^{-20}$}
        \put(-8,23){\scriptsize$10^{-25}$}
        \put(-8,14){\scriptsize$10^{-30}$}
        \put(-8,5){\scriptsize$10^{-35}$}
        \put(18,50){\scriptsize\fboxsep0pt\colorbox{white}{$|\dKK_0{}^0|$}}
        \put(18,44){\scriptsize\fboxsep0pt\colorbox{white}{$|\dKK_2{}^0|$}}
        \put(18,38){\scriptsize\fboxsep0pt\colorbox{white}{$|\dKK_4{}^0|$}}
        \put(18,32){\scriptsize$|\dKK_6{}^0|$}
        \put(18,26){\scriptsize$|\dKK_8{}^0|$}
        \put(18,21){\scriptsize$|\dKK_{10}{}^0|$}
        \put(18,15){\scriptsize$r^{-1.7}$}
        \put(18,9){\scriptsize$r^{-2}$}
      \end{overpic}
      \caption{\footnotesize Falloff rates of $|\dKK_\ell{}^0|$}
      \label{fig:phbrm17KK}
    \end{subfigure}\hspace{.5cm}
    \begin{subfigure}[b]{.5\textwidth}
      \begin{overpic}[width=\textwidth]{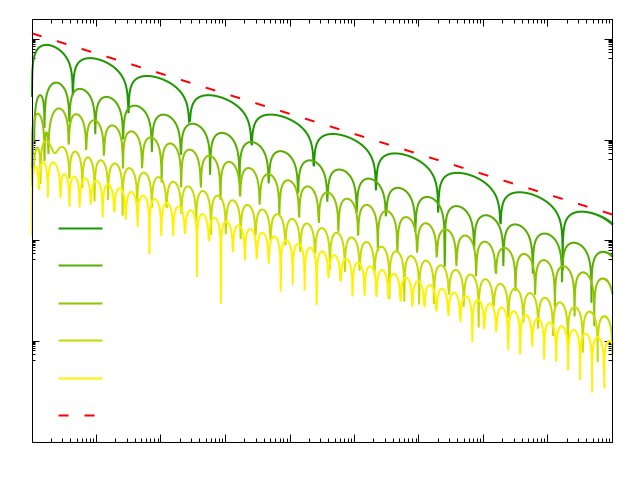}
        \put(4,1){\scriptsize$1$}
        \put(12,1){\scriptsize$10^1$}
        \put(22,1){\scriptsize$10^2$}
        \put(32,1){\scriptsize$10^3$}
        \put(42,1){\scriptsize$10^4$}
        \put(52,1){\scriptsize$10^5$}
        \put(62,1){\scriptsize$10^6$}
        \put(72,1){\scriptsize$10^7$}
        \put(82,1){\scriptsize$10^8$}
        \put(92,1){\scriptsize$10^9$}
        \put(2,67.5){\scriptsize$1$}
        \put(-6,52){\scriptsize$10^{-5}$}
        \put(-8,36){\scriptsize$10^{-10}$}
        \put(-8,21){\scriptsize$10^{-15}$}
        \put(18,38){\scriptsize\fboxsep0pt\colorbox{white}{$|\Re[\dkk_2{}^0]|$}}
        \put(18,32){\scriptsize\fboxsep0pt\colorbox{white}{$|\Re[\dkk_4{}^0]|$}}
        \put(18,26){\scriptsize\fboxsep0pt\colorbox{white}{$|\Re[\dkk_6{}^0]|$}}
        \put(18,21){\scriptsize$|\Re[\dkk_{8}{}^0]|$}
        \put(18,15){\scriptsize$|\Re[\dkk_{10}{}^0]|$}
        \put(18,9){\scriptsize$r^{-1}$}
      \end{overpic}
      \caption{\footnotesize Falloff rates of $|\Re[\dkk_\ell{}^0]|$}
      \label{fig:phbrm17Rekk}
    \end{subfigure}\vskip0.3cm
    \begin{subfigure}[b]{.5\textwidth}
      \begin{overpic}[width=\textwidth]{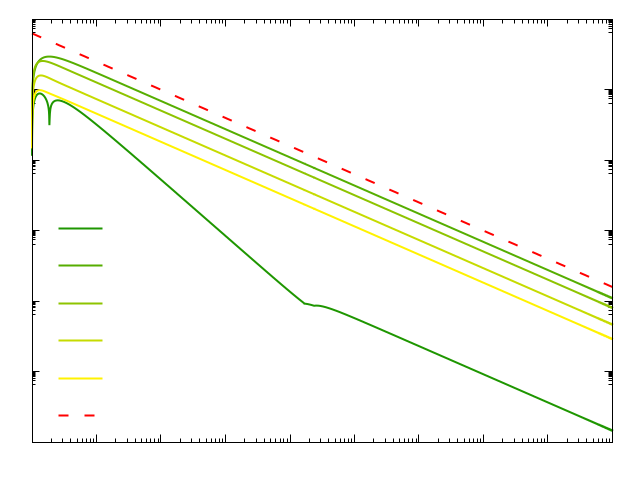}
        \put(4,1){\scriptsize$1$}
        \put(12,1){\scriptsize$10^1$}
        \put(22,1){\scriptsize$10^2$}
        \put(32,1){\scriptsize$10^3$}
        \put(42,1){\scriptsize$10^4$}
        \put(52,1){\scriptsize$10^5$}
        \put(62,1){\scriptsize$10^6$}
        \put(72,1){\scriptsize$10^7$}
        \put(82,1){\scriptsize$10^8$}
        \put(92,1){\scriptsize$10^9$}
        \put(2,71){\scriptsize$1$}
        \put(-6,60){\scriptsize$10^{-5}$}
        \put(-8,49){\scriptsize$10^{-10}$}
        \put(-8,38){\scriptsize$10^{-15}$}
        \put(-8,27){\scriptsize$10^{-20}$}
        \put(-8,16){\scriptsize$10^{-25}$}
        \put(-8,5){\scriptsize$10^{-30}$}
        \put(18,38){\scriptsize\fboxsep0pt\colorbox{white}{$|\Im[\dkk_1{}^0]|$}}
        \put(18,32){\scriptsize\fboxsep0pt\colorbox{white}{$|\Im[\dkk_3{}^0]|$}}
        \put(18,26){\scriptsize$|\Im[\dkk_5{}^0]|$}
        \put(18,21){\scriptsize$|\Im[\dkk_7{}^0]|$}
        \put(18,15){\scriptsize$|\Im[\dkk_9{}^0]|$}
        \put(18,9){\scriptsize$r^{-2}$}
      \end{overpic}
      \caption{\footnotesize Falloff rates of $|\Im[\dkk_\ell{}^0]|$}
      \label{fig:phbrm17Imkk}
    \end{subfigure}\hspace{.5cm}
    \begin{subfigure}[b]{.5\textwidth}
      \begin{overpic}[width=\textwidth]{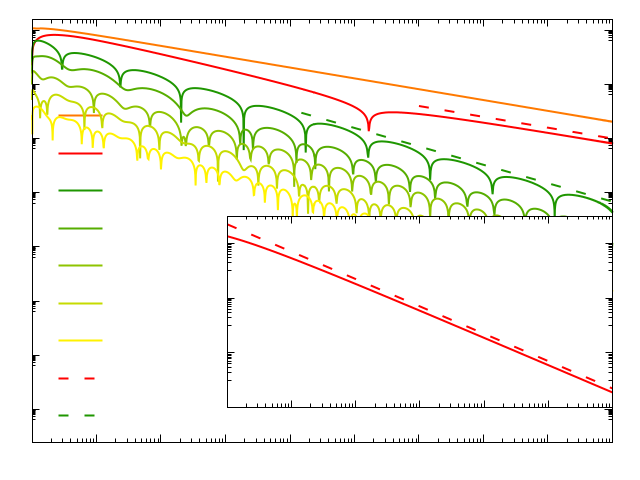}
        \put(4,1){\scriptsize$1$}
        \put(12,1){\scriptsize$10^1$}
        \put(22,1){\scriptsize$10^2$}
        \put(32,1){\scriptsize$10^3$}
        \put(42,1){\scriptsize$10^4$}
        \put(52,1){\scriptsize$10^5$}
        \put(62,1){\scriptsize$10^6$}
        \put(72,1){\scriptsize$10^7$}
        \put(82,1){\scriptsize$10^8$}
        \put(92,1){\scriptsize$10^9$}
        \put(2,69.5){\scriptsize$1$}
        \put(-6,61){\scriptsize$10^{-5}$}
        \put(-8,52.5){\scriptsize$10^{-10}$}
        \put(-8,44){\scriptsize$10^{-15}$}
        \put(-8,36){\scriptsize$10^{-20}$}
        \put(-8,27){\scriptsize$10^{-25}$}
        \put(-8,19){\scriptsize$10^{-30}$}
        \put(-8,10){\scriptsize$10^{-35}$}
        \put(18,56){\scriptsize\fboxsep0pt\colorbox{white}{$|\bgNh_0{}^0-1|$}}
        \put(18,50){\scriptsize\fboxsep0pt\colorbox{white}{$|\dNh_0{}^0|$}}
        \put(18,44){\scriptsize\fboxsep0pt\colorbox{white}{$|\dNh_2{}^0|$}}
        \put(18,38){\scriptsize\fboxsep0pt\colorbox{white}{$|\dNh_4{}^0|$}}
        \put(18,32){\scriptsize$|\dNh_6{}^0|$}
        \put(18,26){\scriptsize$|\dNh_8{}^0|$}
        \put(18,21){\scriptsize$|\dNh_{10}{}^0|$}
        \put(18,15){\scriptsize$r^{-1}$}
        \put(18,9){\scriptsize$r^{-1.7}$}
        \put(32,7){\scriptsize$10^6$}
        \put(42,7){\scriptsize$10^7$}
        \put(52,7){\scriptsize$10^8$}
        \put(62,7){\scriptsize$10^9$}
        \put(72,7){\scriptsize$10^{10}$}
        \put(82,7){\scriptsize$10^{11}$}
        \put(92,7){\scriptsize\fboxsep0pt\colorbox{white}{$10^{12}$}}
      \end{overpic}
      \caption{\footnotesize Falloff rates of $|\dNh_\ell{}^0|$}
      \label{fig:phbrm17Nh}
    \end{subfigure}
    \caption{\footnotesize The falloff rates of the only nontrivial modes of the constrained variables yielded by the parabolic-hyperbolic equations for near-Kerr initial data, and applying \eqref{eq:newscR} with $\alpha=0.7$, are plotted. As $\dNh_0{}^0$ changed its sign close to $r=10^5$ to determine the precise value of the falloff rate, we integrated the parabolic-hyperbolic equations of a longer interval. Note that changing the sign of $\dNh_0{}^0$ does not affect the sign of the ADM mass since $\bgNh_0{}^0-1$, also plotted in \ref{fig:phbrm17Nh}, is always larger, so the sign of $\Nh_0{}^0-1$ does not change. Note that each mode falls off according to the conditions in weak asymptotic flatness.}
    \label{fig:phbrm17}
  \end{figure}

	In order to demonstrate that weakly asymptotically flat initial data can also be produced by integrating the parabolic-hyperbolic equations in the near-Kerr case, we determined the falloff rates of various modes of the constrained variables using \eqref{eq:newscR} with $\alpha=0.7$. These are depicted in Fig.\,\ref{fig:phbrm17}. The monopole mode $\dKK_0{}^0$ falls off, as expected, with the rate $\mathcal{O}(r^{-1.7})$, whereas all the other $\dKK_\ell{}^0|_{\ell>0}$ modes decay with the rate $\mathcal{O}(r^{-2})$. The falloff behavior of the $\dkk_\ell{}^0$ modes follows the rule observed in the $\alpha=1$ case. As $\dNh_0{}^0$ changed its sign close to $r=10^5$ to determine the falloff rate's precise value, we integrated the parabolic-hyperbolic equations on a longer interval. The observed falloff rate of $\dNh_0{}^0$ is, as expected, $\mathcal{O}(r^{-1})$. Note that changing the sign of $\dNh_0{}^0$ does not affect the sign of the ADM mass since $\bgNh_0{}^0-1$, also plotted in Fig. \ref{fig:phbrm17Nh} (see top curve), is always larger, so the sign of $\Nh_0{}^0-1$ does not change. Note also that all the higher $\ell$ modes of $\dNh_\ell{}^0$ fall off with the rate $\mathcal{O}(r^{-1.7})$. It is important to emphasize that all the reported falloff rates are compatible with conditions of weak asymptotic flatness in \eqref{eq:weak-asymp}.

	As the plots in Figs.\,\ref{fig:phbrm2} and \ref{fig:phbrm17} demonstrate, the choice we made for the monopole part of $\kkappa$ imposing \eqref{eq:newscR} allowed us to produce both weakly and strongly asymptotically flat near-Kerr initial data configurations. Another preferable consequence of this choice was that it left intact the principal part of the parabolic-hyperbolic equations. Although the implicit averaging used to determine the monopole part of $\kkappa$ may raise doubts about the nature of the system, the results  in \cite{Racz:2015mfa} on the principal symbol of the equations, together with the observed convergence properties, support our belief in the well posedness of the corresponding initial value problem.

  \subsection{Applying the new method to the algebraic-hyperbolic system}
  \label{sec:ahm}

	In this section, using the algebraic-hyperbolic form of the constraint equations, we impose a condition only on the monopole part of the lapse $\Nh$. In contrast, all the higher $\ell$ modes $\Nh$ retain their background form. In particular, $\Nh_0{}^0$ is subject to the ODE derived from \eqref{eq:phN}, which is as follows:
	\begin{align}\label{eq:ahmODE}
		\frac{\mathrm{d}\Nh_0{}^0}{\mathrm{d}r} = \big[(\partial_r\Nh)+\Kstar{}^{-1}\big\{\widetilde{\mathcal{B}}-\mathcal{B}\big\}\Nh^3\big]_0{}^0\,.
	\end{align}
	Here $(\partial_r\Nh)$ denotes the right-hand side of the equation obtained by solving \eqref{eq:phN} for $\partial_r\Nh$, whereas
	\begin{equation}
		 \Nh=\Nh_0{}^0+[\bgNh-\bgNh_0{}^0]\,,
	\end{equation}
	i.e., $\Nh$ differs from $\bgNh$ only in its monopole part, implying $\eth\Nh=\eth\bgNh$.	Note also that the monopole part of the right-hand side \eqref{eq:ahmODE} is taken, $\widetilde{\mathcal{B}}$ is obtained from $\mathcal{B}$, as given in \eqref{eq:phB}, by replacing\,\footnote{Note that this step is exactly the same procedure as in Sec. \ref{sec:phm}.} $\kkappa$ with
	\begin{equation}
		\alpha\left[\mathscr{R}\,\KK\right]_0{}^0+ [\bgkkappa-\bgkkappa_0{}^0]\,,\quad {\rm with} \ \mathscr{R} = \bgkkappa/\bgKK\,.
	\end{equation}
	Note that this choice of $\kkappa$ only affects the monopole part of $\Nh$ via \eqref{eq:ahmODE}. Note also that the true physical value of $\kkappa$ is determined (as it should be) by the algebraic form of the Hamiltonian constraint \eqref{eq:ahkappa}. As we will see below, the use of the positive parameter $\alpha$ in \eqref{eq:ahmODE} allows fine-tuning of the decay rate of the monopole part of $\KK$. Note also that \eqref{eq:ahmODE} is able to reproduce the Kerr limit by choosing $\alpha=1$ and trivial excitations for the fields $\KK,\kk$ at $r=1$. It is also straightforward to derive the spherical limit by applying the relations $\Nh=\bgNh=\Nh_0{}^0$.

\subsubsection{Outline in the spherically symmetric setup}

	Note that there is not much earned by restricting considerations to the spherically symmetric case as the algebraic-hyperbolic system is essentially the same as it was in the parabolic-hyperbolic case. More concretely, the algebraic form of the Hamiltonian constraint is nothing but solving \eqref{eq:phmNsph} for $\kkappa$ or $2\,\kkappa\,\KK$, whereas Eq. \eqref{eq:phmKsph} for $\KK$ remains intact. Accordingly, the corresponding spherical symmetric system with Schwarzschild background reads as
	\begin{align}
		2\,\kkappa\,\KK&=-\tfrac{2}{r^2}-\tfrac{1}{2}\KK^2+\tfrac{2}{r^2\Nh^2}-\tfrac{4}{r\Nh^3}\tfrac{\mathrm{d}\Nh}{\mathrm{d}r}\,,\label{eq:ahmkappa}\\ \frac{\mathrm{d}\KK}{\mathrm{d}r}&=\tfrac{2}{r}\left(\kkappa-\tfrac{1}{2}\KK\right)\,.\label{eq:ahmKK}
	\end{align}
	Note that by eliminating the $r$ derivative of $\Nh$ from \eqref{eq:ahmkappa} by applying the spherical symmetric form of \eqref{eq:ahmODE} one recovers \eqref{eq:newscR}. In light of this observation, the analytic solutions to  \eqref{eq:ahmKK} must possess the form \eqref{eq:phmKKsol} with asymptotic behavior as given in \eqref{eq:phmKKasym}. Accordingly, the parameter $\alpha$, applied in \eqref{eq:ahmODE}, allows a straightforward fine-tuning of the asymptotic falloff rate of $\KK$. As previously, in the spherically symmetric parabolic-hyperbolic case, to guarantee $\KK$ to fit the requirements in weak asymptotic flatness, the inequality $\alpha>1/2$ must hold. In the $\alpha=1$ case, the falloff rate in \eqref{eq:phmKKasym} allows the spherically symmetric solution to be strongly asymptotically flat.

\subsubsection{Strongly asymptotically flat numerical solution}

\begin{figure}
    \begin{subfigure}[b]{.5\textwidth}
      \begin{overpic}[width=\textwidth]{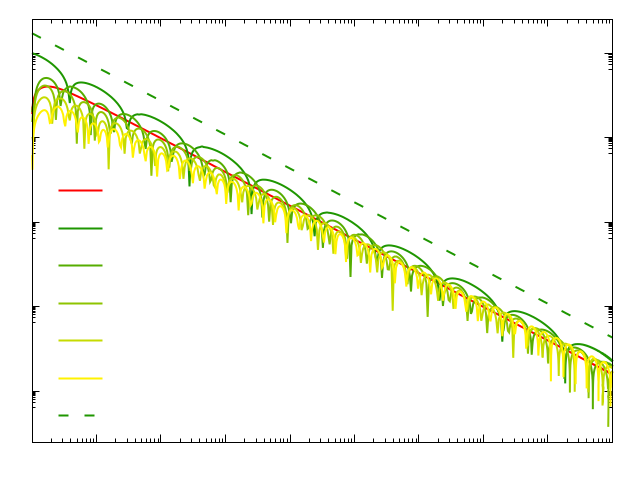}
        \put(4,1){\scriptsize$1$}
        \put(12,1){\scriptsize$10^1$}
        \put(22,1){\scriptsize$10^2$}
        \put(32,1){\scriptsize$10^3$}
        \put(42,1){\scriptsize$10^4$}
        \put(52,1){\scriptsize$10^5$}
        \put(62,1){\scriptsize$10^6$}
        \put(72,1){\scriptsize$10^7$}
        \put(82,1){\scriptsize$10^8$}
        \put(92,1){\scriptsize$10^9$}
        \put(2,66){\scriptsize$1$}
        \put(-6,52){\scriptsize$10^{-5}$}
        \put(-8,39){\scriptsize$10^{-10}$}
        \put(-8,26.5){\scriptsize$10^{-15}$}
        \put(-8,13){\scriptsize$10^{-20}$}
        \put(18,44){\scriptsize\fboxsep0pt\colorbox{white}{$|\dKK_0{}^0|$}}
        \put(18,38){\scriptsize\fboxsep0pt\colorbox{white}{$|\dKK_2{}^0|$}}
        \put(18,32){\scriptsize$|\dKK_4{}^0|$}
        \put(18,26){\scriptsize$|\dKK_6{}^0|$}
        \put(18,21){\scriptsize$|\dKK_8{}^0|$}
        \put(18,15){\scriptsize$|\dKK_{10}{}^0|$}
        \put(18,9){\scriptsize$r^{-2}$}
      \end{overpic}
      \caption{\footnotesize Falloff rates of $|\dKK_\ell{}^0|$}
      \label{fig:ahmrm2KK}
    \end{subfigure}\hspace{.5cm}
    \begin{subfigure}[b]{.5\textwidth}
      \begin{overpic}[width=\textwidth]{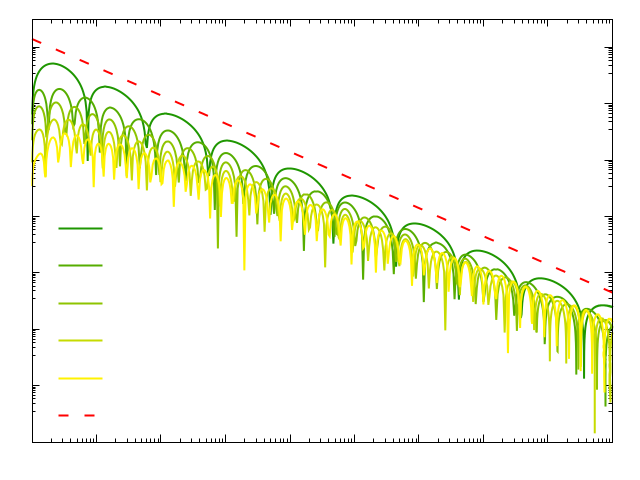}
        \put(4,1){\scriptsize$1$}
        \put(12,1){\scriptsize$10^1$}
        \put(22,1){\scriptsize$10^2$}
        \put(32,1){\scriptsize$10^3$}
        \put(42,1){\scriptsize$10^4$}
        \put(52,1){\scriptsize$10^5$}
        \put(62,1){\scriptsize$10^6$}
        \put(72,1){\scriptsize$10^7$}
        \put(82,1){\scriptsize$10^8$}
        \put(92,1){\scriptsize$10^9$}
        \put(2,66.5){\scriptsize$1$}
        \put(-6,58){\scriptsize$10^{-2}$}
        \put(-6,49){\scriptsize$10^{-4}$}
        \put(-6,40){\scriptsize$10^{-6}$}
        \put(-6,31){\scriptsize$10^{-8}$}
        \put(-8,22){\scriptsize$10^{-10}$}
        \put(-8,14){\scriptsize$10^{-12}$}
        \put(-8,5){\scriptsize$10^{-14}$}
        \put(18,38){\scriptsize\fboxsep0pt\colorbox{white}{$|\Re[\dkk_2{}^0]|$}}
        \put(18,32){\scriptsize\fboxsep0pt\colorbox{white}{$|\Re[\dkk_4{}^0]|$}}
        \put(18,26){\scriptsize$|\Re[\dkk_6{}^0]|$}
        \put(18,21){\scriptsize$|\Re[\dkk_{8}{}^0]|$}
        \put(18,15){\scriptsize$|\Re[\dkk_{10}{}^0]|$}
        \put(18,9){\scriptsize$r^{-1}$}
      \end{overpic}
      \caption{\footnotesize Falloff rates of $|\Re[\dkk_\ell{}^0]|$}
      \label{fig:ahmrm2rekk}
    \end{subfigure}\\
    \begin{subfigure}[b]{.5\textwidth}
      \begin{overpic}[width=\textwidth]{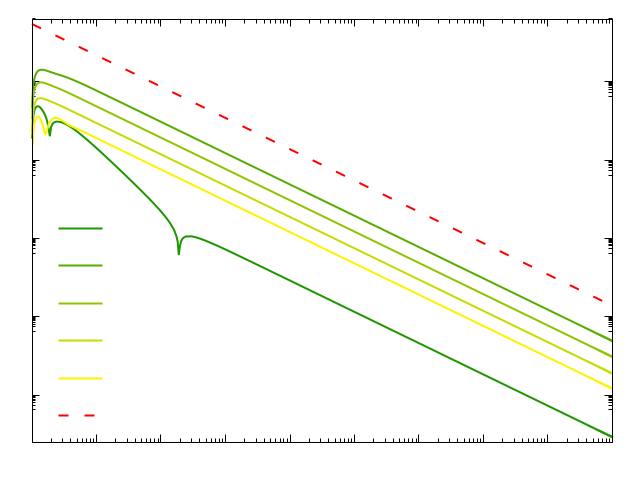}
        \put(4,1){\scriptsize$1$}
        \put(12,1){\scriptsize$10^1$}
        \put(22,1){\scriptsize$10^2$}
        \put(32,1){\scriptsize$10^3$}
        \put(42,1){\scriptsize$10^4$}
        \put(52,1){\scriptsize$10^5$}
        \put(62,1){\scriptsize$10^6$}
        \put(72,1){\scriptsize$10^7$}
        \put(82,1){\scriptsize$10^8$}
        \put(92,1){\scriptsize$10^9$}
        \put(-6,61){\scriptsize$10^{-5}$}
        \put(-8,48.5){\scriptsize$10^{-10}$}
        \put(-8,36.5){\scriptsize$10^{-15}$}
        \put(-8,24.5){\scriptsize$10^{-20}$}
        \put(-8,12){\scriptsize$10^{-25}$}
        \put(18,38){\scriptsize\fboxsep0pt\colorbox{white}{$|\Im[\dkk_1{}^0]|$}}
        \put(18,32){\scriptsize\fboxsep0pt\colorbox{white}{$|\Im[\dkk_3{}^0]|$}}
        \put(18,26){\scriptsize$|\Im[\dkk_5{}^0]|$}
        \put(18,21){\scriptsize$|\Im[\dkk_7{}^0]|$}
        \put(18,15){\scriptsize$|\Im[\dkk_9{}^0]|$}
        \put(18,9){\scriptsize$r^{-2}$}
      \end{overpic}
     	 \caption{\footnotesize Falloff rates of $|\Im[\dkk_\ell{}^0]|$}
     	 \label{fig:ahmrm2imkk}
      \end{subfigure}\hspace{.5cm}
	  \begin{subfigure}[b]{.5\textwidth}
	  	\begin{overpic}[width=\textwidth]{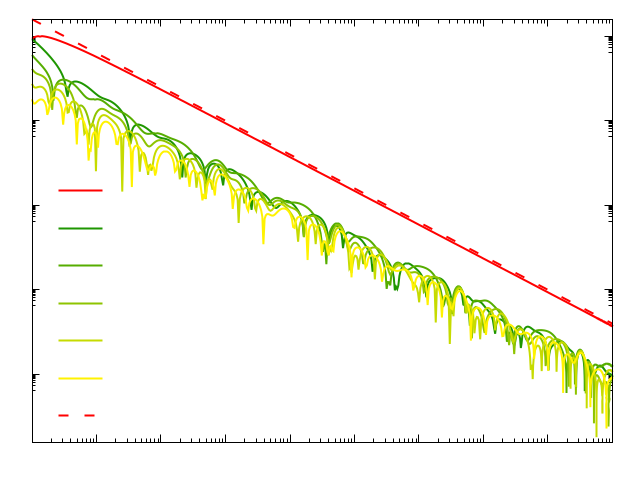}
	  		\put(4,1){\scriptsize$1$}
	  		\put(12,1){\scriptsize$10^1$}
	  		\put(22,1){\scriptsize$10^2$}
	  		\put(32,1){\scriptsize$10^3$}
	  		\put(42,1){\scriptsize$10^4$}
	  		\put(52,1){\scriptsize$10^5$}
	  		\put(62,1){\scriptsize$10^6$}
	  		\put(72,1){\scriptsize$10^7$}
	  		\put(82,1){\scriptsize$10^8$}
	  		\put(92,1){\scriptsize$10^9$}
	  		\put(-6,55.5){\scriptsize$10^{-5}$}
	  		\put(-8,42){\scriptsize$10^{-10}$}
	  		\put(-8,29){\scriptsize$10^{-15}$}
	  		\put(-8,16){\scriptsize$10^{-20}$}
	  		\put(2,68.5){\scriptsize$1$}
        \put(18,44){\scriptsize\fboxsep0pt\colorbox{white}{$|\kkappa_0{}^0|$}}
        \put(18,38){\scriptsize\fboxsep0pt\colorbox{white}{$|\kkappa_2{}^0|$}}
        \put(18,32){\scriptsize$|\kkappa_4{}^0|$}
        \put(18,26){\scriptsize$|\kkappa_6{}^0|$}
        \put(18,21){\scriptsize$|\kkappa_8{}^0|$}
        \put(18,15){\scriptsize$|\kkappa_{10}{}^0|$}
	  		\put(18,9){\scriptsize$r^{-2}$}
	  	\end{overpic}
	  	\caption{\footnotesize Falloff rates of $|\kkappa_\ell{}^0|$}
	  	\label{fig:ahmrm2imkkappa}
	  \end{subfigure}
    \caption{\footnotesize The falloff rates of the only nontrivial modes of the constrained variables yielded by the algebraic-hyperbolic equations for near-Kerr initial data, and applying  \eqref{eq:ahmODE} with $\alpha=1$, are plotted. Note that each mode falls off in accordance with the conditions in strong asymptotic flatness.}
    \label{fig:ahmrm2}
  \end{figure}
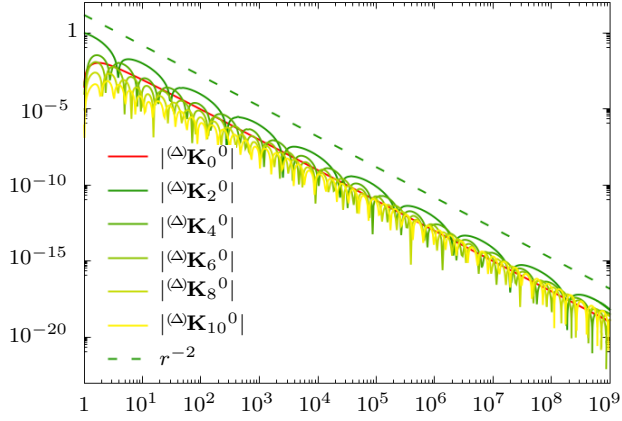
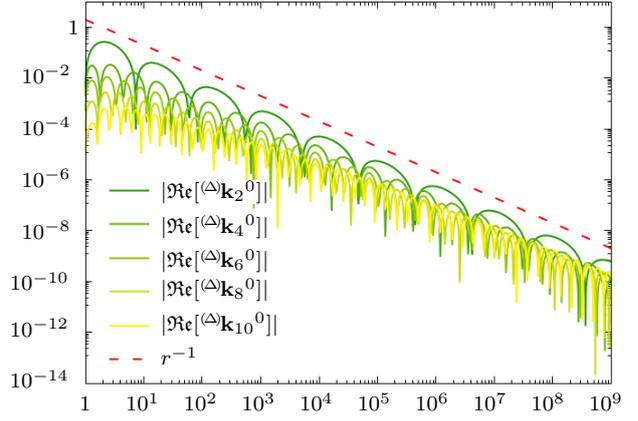
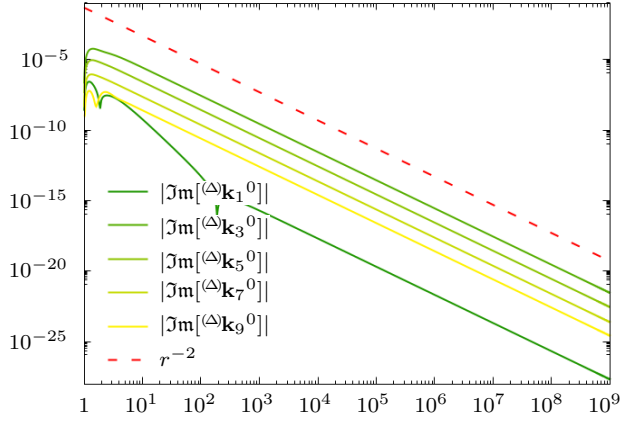
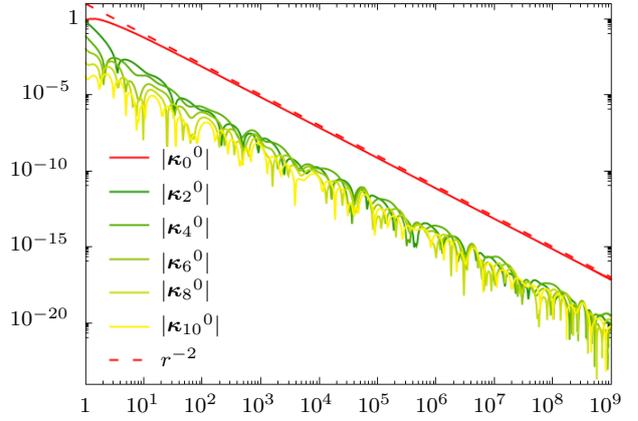

	In Fig.\,\ref{fig:ahmrm2} the falloff rates of the constraint fields $\dKK$, $\dkk$, and $\kkappa$, yielded by the integration of the algebraic-hyperbolic equations, \eqref{eq:ahK}-\eqref{eq:ahk}, for near-Kerr initial data, and applying \eqref{eq:ahmODE} with $\alpha=1$, are depicted. In this case each of the $\dKK_\ell{}^0$ modes falls off as $\mathcal{O}(r^{-2})$, $\Re[\dkk_\ell{}^0]$ as $\mathcal{O}(r^{-1})$ with $\Im[\dkk_\ell{}^0]$ going as $\mathcal{O}(r^{-2})$. Note that using the algebraic-hyperbolic equations, we had to perform several divisions by $\KK$, which, because of the uniform decay rate of all the $\dKK_\ell{}^0$ modes, $\ell=0,2,4,6,8,10$, required a careful application of the division scheme based on the Neumann series expansion. Note also the falloff of the modes of $\dkk$ follow the same pattern observed in Sec. \ref{sec:phm}. Since all modes of $\kkappa$ also fall off with the rate $\mathcal{O}(r^{-2})$, the initial data obtained is asymptotically flat in the strong sense. Note that although $\Nh_0{}^0$ also deviates from its background form, it falls at the rate $\mathcal{O}(r^{-1})$ to $1$ as expected, so we do not include its graph here.

\subsubsection{Weakly asymptotically flat numerical solution}

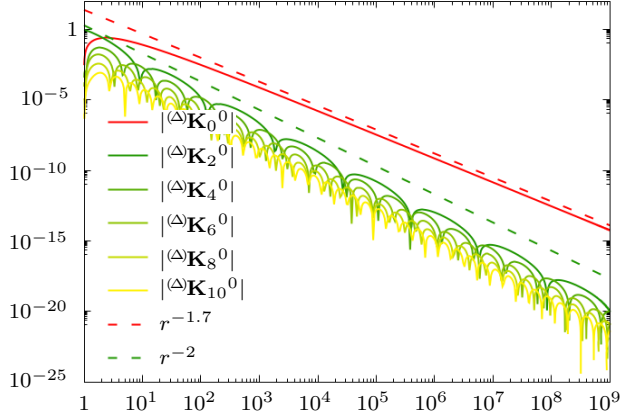
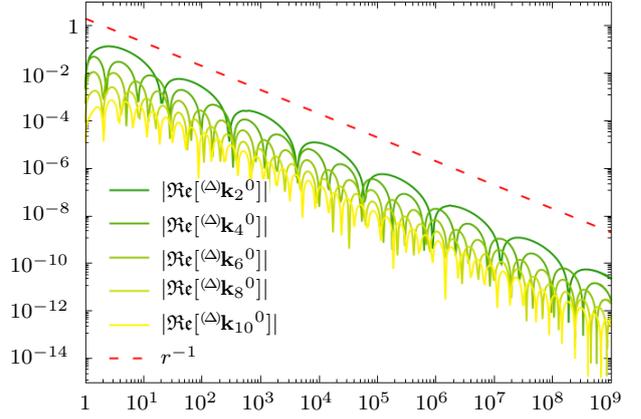
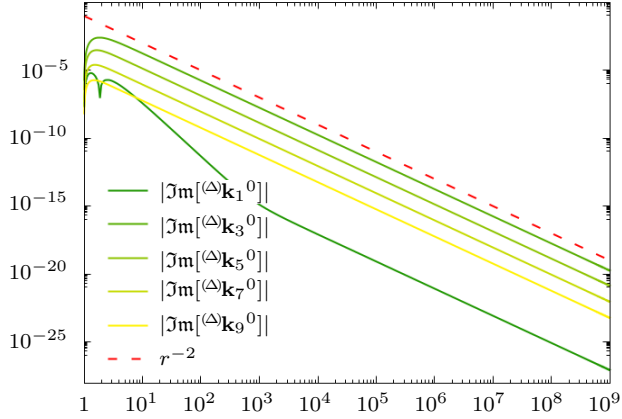
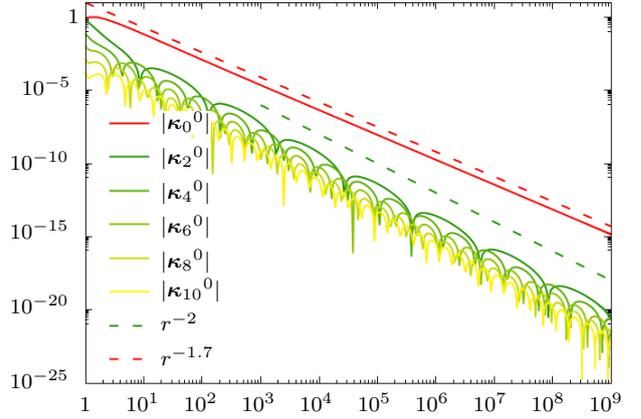
\begin{figure}
    \begin{subfigure}[b]{.5\textwidth}
      \begin{overpic}[width=\textwidth]{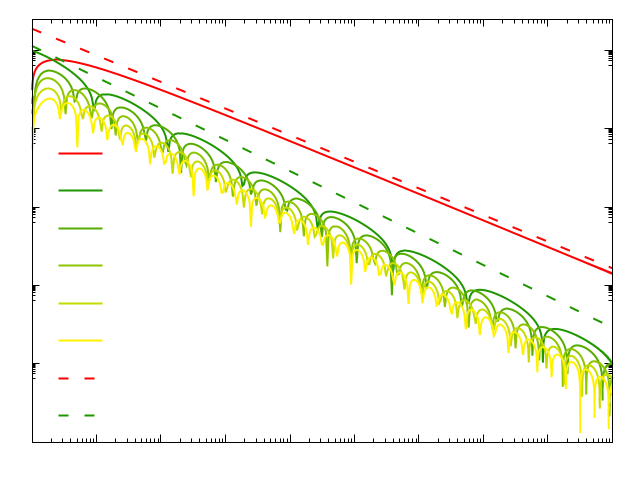}
        \put(4,1){\scriptsize$1$}
        \put(12,1){\scriptsize$10^1$}
        \put(22,1){\scriptsize$10^2$}
        \put(32,1){\scriptsize$10^3$}
        \put(42,1){\scriptsize$10^4$}
        \put(52,1){\scriptsize$10^5$}
        \put(62,1){\scriptsize$10^6$}
        \put(72,1){\scriptsize$10^7$}
        \put(82,1){\scriptsize$10^8$}
        \put(92,1){\scriptsize$10^9$}
        \put(2,66){\scriptsize$1$}
        \put(-6,54){\scriptsize$10^{-5}$}
        \put(-8,41){\scriptsize$10^{-10}$}
        \put(-8,29){\scriptsize$10^{-15}$}
        \put(-8,17){\scriptsize$10^{-20}$}
        \put(-8,5){\scriptsize$10^{-25}$}
        \put(18,50){\scriptsize\fboxsep0pt\colorbox{white}{$|\dKK_0{}^0|$}}
        \put(18,44){\scriptsize\fboxsep0pt\colorbox{white}{$|\dKK_2{}^0|$}}
        \put(18,38){\scriptsize\fboxsep0pt\colorbox{white}{$|\dKK_4{}^0|$}}
        \put(18,32){\scriptsize$|\dKK_6{}^0|$}
        \put(18,26){\scriptsize$|\dKK_8{}^0|$}
        \put(18,21){\scriptsize$|\dKK_{10}{}^0|$}
        \put(18,15){\scriptsize$r^{-1.7}$}
        \put(18,9){\scriptsize$r^{-2}$}
      \end{overpic}
      \caption{\footnotesize Falloff rates of $|\dKK_\ell{}^0|$}
      \label{fig:ahmrm17KK}
    \end{subfigure}\hspace{.5cm}
    \begin{subfigure}[b]{.5\textwidth}
      \begin{overpic}[width=\textwidth]{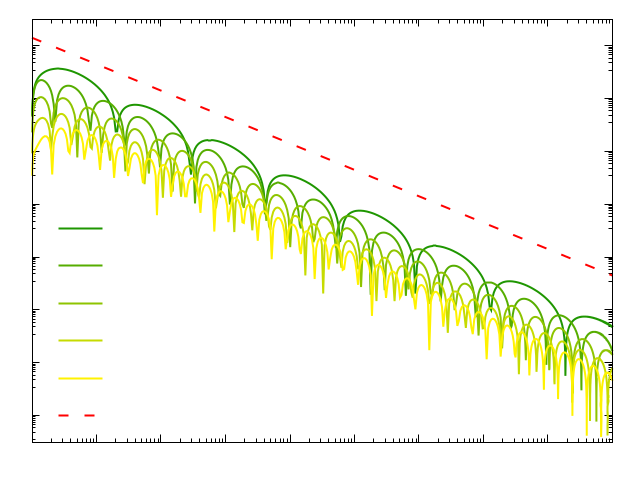}
        \put(4,1){\scriptsize$1$}
        \put(12,1){\scriptsize$10^1$}
        \put(22,1){\scriptsize$10^2$}
        \put(32,1){\scriptsize$10^3$}
        \put(42,1){\scriptsize$10^4$}
        \put(52,1){\scriptsize$10^5$}
        \put(62,1){\scriptsize$10^6$}
        \put(72,1){\scriptsize$10^7$}
        \put(82,1){\scriptsize$10^8$}
        \put(92,1){\scriptsize$10^9$}
        \put(2,66.5){\scriptsize$1$}
        \put(-6,58){\scriptsize$10^{-2}$}
        \put(-6,50){\scriptsize$10^{-4}$}
        \put(-6,42){\scriptsize$10^{-6}$}
        \put(-6,34){\scriptsize$10^{-8}$}
        \put(-8,25){\scriptsize$10^{-10}$}
        \put(-8,17){\scriptsize$10^{-12}$}
        \put(-8,9){\scriptsize$10^{-14}$}
        \put(18,38){\scriptsize\fboxsep0pt\colorbox{white}{$|\Re[\dkk_2{}^0]|$}}
        \put(18,32){\scriptsize\fboxsep0pt\colorbox{white}{$|\Re[\dkk_4{}^0]|$}}
        \put(18,26){\scriptsize$|\Re[\dkk_6{}^0]|$}
        \put(18,21){\scriptsize$|\Re[\dkk_{8}{}^0]|$}
        \put(18,15){\scriptsize$|\Re[\dkk_{10}{}^0]|$}
        \put(18,9){\scriptsize$r^{-1}$}
      \end{overpic}
      \caption{\footnotesize Falloff rates of $|\Re[\dkk_\ell{}^0]|$}
      \label{fig:ahmrm17Rekk}
    \end{subfigure}\\
    \begin{subfigure}[b]{.5\textwidth}
      \begin{overpic}[width=\textwidth]{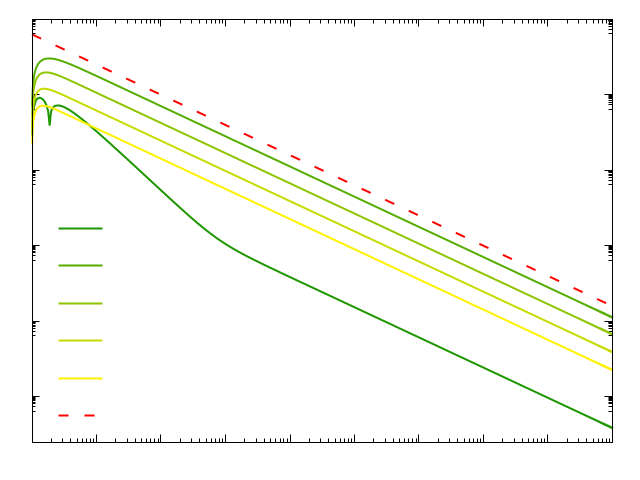}
        \put(4,1){\scriptsize$1$}
        \put(12,1){\scriptsize$10^1$}
        \put(22,1){\scriptsize$10^2$}
        \put(32,1){\scriptsize$10^3$}
        \put(42,1){\scriptsize$10^4$}
        \put(52,1){\scriptsize$10^5$}
        \put(62,1){\scriptsize$10^6$}
        \put(72,1){\scriptsize$10^7$}
        \put(82,1){\scriptsize$10^8$}
        \put(92,1){\scriptsize$10^9$}
        \put(-6,59){\scriptsize$10^{-5}$}
        \put(-8,47.5){\scriptsize$10^{-10}$}
        \put(-8,36){\scriptsize$10^{-15}$}
        \put(-8,24){\scriptsize$10^{-20}$}
        \put(-8,12){\scriptsize$10^{-25}$}
        \put(18,38){\scriptsize\fboxsep0pt\colorbox{white}{$|\Im[\dkk_1{}^0]|$}}
        \put(18,32){\scriptsize\fboxsep0pt\colorbox{white}{$|\Im[\dkk_3{}^0]|$}}
        \put(18,26){\scriptsize$|\Im[\dkk_5{}^0]|$}
        \put(18,21){\scriptsize$|\Im[\dkk_7{}^0]|$}
        \put(18,15){\scriptsize$|\Im[\dkk_9{}^0]|$}
        \put(18,9){\scriptsize$r^{-2}$}
      \end{overpic}
      \caption{\footnotesize Falloff rates of $|\Im[\dkk_\ell{}^0]|$}
      \label{fig:ahmrm17Imkk}
    \end{subfigure}\hspace{.5cm}
	\begin{subfigure}[b]{.5\textwidth}
		\begin{overpic}[width=\textwidth]{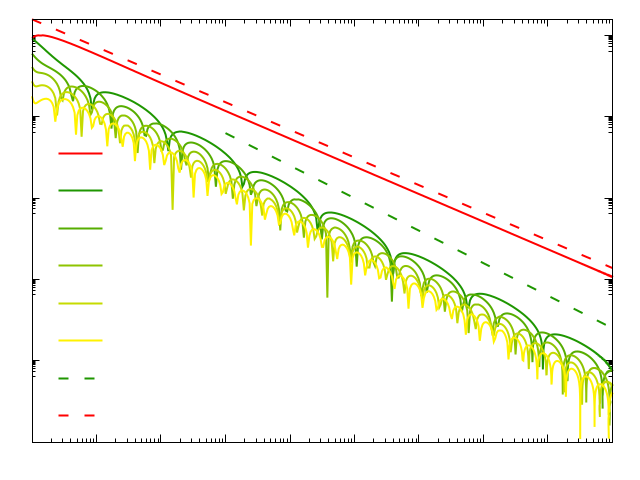}
			\put(4,1){\scriptsize$1$}
			\put(12,1){\scriptsize$10^1$}
			\put(22,1){\scriptsize$10^2$}
			\put(32,1){\scriptsize$10^3$}
			\put(42,1){\scriptsize$10^4$}
			\put(52,1){\scriptsize$10^5$}
			\put(62,1){\scriptsize$10^6$}
			\put(72,1){\scriptsize$10^7$}
			\put(82,1){\scriptsize$10^8$}
			\put(92,1){\scriptsize$10^9$}
      \put(2,68.5){\scriptsize$1$}
			\put(-6,56){\scriptsize$10^{-5}$}
			\put(-8,43){\scriptsize$10^{-10}$}
			\put(-8,31){\scriptsize$10^{-15}$}
			\put(-8,18){\scriptsize$10^{-20}$}
			\put(-8,5){\scriptsize$10^{-25}$}
      \put(18,50){\scriptsize\fboxsep0pt\colorbox{white}{$|\kkappa_0{}^0|$}}
      \put(18,44){\scriptsize\fboxsep0pt\colorbox{white}{$|\kkappa_2{}^0|$}}
      \put(18,38){\scriptsize\fboxsep0pt\colorbox{white}{$|\kkappa_4{}^0|$}}
      \put(18,32){\scriptsize$|\kkappa_6{}^0|$}
      \put(18,26){\scriptsize$|\kkappa_8{}^0|$}
      \put(18,21){\scriptsize$|\kkappa_{10}{}^0|$}
      \put(18,15){\scriptsize$r^{-2}$}
      \put(18,9){\scriptsize$r^{-1.7}$}
		\end{overpic}
		\caption{\footnotesize Falloff rates of $|\kkappa_\ell{}^0|$}
		\label{fig:ahmrm17Imkkappa}
	\end{subfigure}
    \caption{\footnotesize  The falloff rates of the only nontrivial modes of the constrained variables yielded by integrating the algebraic-hyperbolic equations for near-Kerr initial data, and applying \eqref{eq:ahmODE} with $\alpha=0.7$, are plotted. Note that each mode falls off according to the conditions in weak asymptotic flatness. }
    \label{fig:ahmrm17}
  \end{figure}

	In order to demonstrate that weakly asymptotically flat initial data can also be produced by integrating the algebraic-hyperbolic equations in the near-Kerr case, we determined the falloff rates of various modes of the constrained variables using \eqref{eq:ahmODE} with $\alpha=0.7$. These are depicted in Fig.\,\ref{fig:ahmrm17}. The monopole mode $\dKK_0{}^0$ falls off, as expected, with the rate $\mathcal{O}(r^{-1.7})$, whereas all the other $\dKK_\ell{}^0|_{\ell>0}$ modes decay---likewise it happened in the strictly near-Kerr case in Sec. \ref{sec:strictscheme}---somewhat slower than $\mathcal{O}(r^{-2.2})$ but noticeably faster than $\mathcal{O}(r^{-2})$. As for the asymptotic behavior of the modes $\Re[\dkk_\ell{}^0]$ also somewhat slower than $\mathcal{O}(r^{-1.2})$ but noticeably faster than $\mathcal{O}(r^{-1})$ falloff rate is observed. The falloff rate for the  $\Im[\dkk_\ell{}^0]$ modes is as usual $\mathcal{O}(r^{-2})$. Similar to $\dKK$, $\kkappa_0{}^0$ decays as $\mathcal{O}(r^{-1.7})$, while slightly faster than $\mathcal{O}(r^{-2})$ decay rates are observed for $\ell>0$ modes of $\kkappa$. It is important to emphasize that all the observed falloff rates are compatible with conditions of weak asymptotic flatness in \eqref{eq:weak-asymp}, i.e., the resulted initial data is  asymptotically flat in the weaker sense.

	The plots in Figs.\,\ref{fig:ahmrm2} and \ref{fig:ahmrm17} demonstrate that the choice we made for the monopole part of $\Nh$ imposing  \eqref{eq:ahmODE} allowed us to produce both weakly and strongly asymptotically flat near-Kerr initial data configurations by integrating the algebraic-hyperbolic system. As it was emphasized several times, restricting the monopole part of $\Nh$ using \eqref{eq:ahmODE} has the preferable consequence that it does not affect the principal part of the algebraic-hyperbolic equations. Note also that the parameter $\alpha$ in \eqref{eq:ahmODE} could be used to control the falloff rate of $\dKK_0{}^0$ in the desired way.

	As expected, the solutions to the evolutionary form of the constraints significantly depend on whether \eqref{eq:newscR} or  \eqref{eq:ahmODE} is applied. To demonstrate this, it is rewarding to compare the falloff behavior of the corresponding variables yielded by using the same initial excitation as specified in \eqref{eq:initial-data}. The falloff rates of the $\ell>0, m=0$ modes of the variables $\KK,\Re[\kk],\Im[\kk],\Nh,\kkappa$ are collected in Table\,\ref{table:fall-off} relevant for the weakly asymptotically flat solutions to the parabolic-hyperbolic and algebraic-hyperbolic systems corresponding to the choice $\alpha=0.7$.
	\begin{table}[H]
		\centering  \hskip-.15cm
		\begin{tabular}{|c||c|c|c|c|c|}
			\hline $\alpha=0.7$; $\ell>0, m=0$ & $\KK_\ell{}^0$ &  $\Re[\kk_\ell{}^0]$   & $\Im[\kk_\ell{}^0]$  & $\Nh_\ell{}^0$ & $\kkappa_\ell{}^0$ \\ \hline
		Parabolic-hyperbolic	& $\mathcal{O}(r^{-2})$ & $\mathcal{O}(r^{-1})$ & $\mathcal{O}(r^{-2})$ & $\mathcal{O}(r^{-1.7})$ & $\mathcal{O}(r^{-\ell-2})$ \\
			\hline
		Algebraic-hyperbolic	& $\mathcal{O}(r^{-2.2+\epsilon})$ & $\mathcal{O}(r^{-1.2+\epsilon})$ & $\mathcal{O}(r^{-2})$ & $\mathcal{O}(r^{-\ell-1})$ & $\mathcal{O}(r^{-2.2+\epsilon})$ \\
		\hline
		\end{tabular}
		\caption{\small The falloff rates of the $\ell>0,m=0$ modes of $\KK$, $\Re[\kk]$, $\Im[\kk]$, $\Nh$, $\kkappa$ for the weakly asymptotically flat solutions to the parabolic-hyperbolic and algebraic-hyperbolic systems with $\alpha=0.7$, and for some small $\epsilon>0$.}\label{table:fall-off}
	\end{table}

  \section{Summary}
  \label{sec:summary}

	As mentioned earlier, one of the preferable aspects of using the elliptic (or conformal) method is that one may set up a boundary value problem by implementing suitable falloff requirements at infinity. This cannot be done while solving the evolutionary form of the constraints. Nevertheless, in the latter case, a delicate compensation arises. While one has to fix once and for all throughout $\Sigma$ each of the freely specifiable variables in advance of solving the elliptic problem, one has the freedom of specifying in the interim of the ``time integration'' process the freely specifiable fields while solving either of the evolutionary forms of the constraints.
  The only limitation is that this in-flight setting of the freely specifiable fields should preserve the well posedness of the system. Our novel proposal to relate only the monopole part of a freely specifiable variable and a constrained field leaves the principal symbol of the system intact. Due to the nonlocality, introduced by the averaging in the proposed treatment of the monopole part, the well posedness of the system is not self-evident. Nevertheless, since the principal parts of the evolutionary systems remain intact, the observed convergence rates suggest that the proposed method leads to well-posed problems in both evolutionary formulations of the constraints. Note, however, that a rigorous verification of this conjecture is beyond the scope of the present paper.

	\medskip

	Concerning the question raised by the title and some others raised in the Introduction, the results reported in this paper allow us to answer them in the following way. Yes, it is possible to construct asymptotically flat initial data by applying either of the evolutionary forms of the constraints. Above this, we could introduce a novel method that allows us to explore the asymptotic behavior of initial datasets located in a neighborhood of the data that can be deduced on a Kerr-Schild time slice of a rotating Kerr black hole spacetime. In applying our new method, we modify only the monopole part of the freely specifiable variables $\kkappa$ or $\Nh$ in the parabolic-hyperbolic or algebraic-hyperbolic form of the constraint, respectively. On the one hand, this left the principal parts of the evolutionary forms of the constraints intact. Thus, the well posedness of evolutionary systems seems to be plausible in line with the results in \cite{Racz:2015mfa}.

	On the other hand,  our proposal can be applied to the alternative evolutionary systems on an equal footing. In contrast, the method proposed in \cite{Beyer:2020zlo} applies only to the parabolic-hyperbolic system with the additional cost of modifying the principal part of the pertinent PDEs. Another favorable aspect of the novel method proposed in this paper, which applies to both of the evolutionary forms of the constraints, is that we have direct control of the falloff rate of the monopole part of the constrained variable $\KK$. This allows us to generate weakly and strongly asymptotically flat initial data configurations with any desired falloff rate. The viability of the proposed methods was demonstrated by integrating both of the alternative evolutionary forms of the constraint equations numerically. As the applied initial data deformations at $r=1$ were significant, the time evolution of the yielded near-Kerr initial data configurations could be suitable to study the gravitational wave output of highly excited Kerr black holes.

	\medskip

	To indicate the notable differences between applying \eqref{eq:newscR} and the choice made in \cite{Beyer:2020zlo}, it is rewarding to compare the solutions to the relevant evolutionary forms of the constraint equations with setting $\alpha=1.5$. As for  $\mathcal{R}=\alpha\cdot\mathscr{R}=\alpha\cdot \bgkkappa_0{}^0/\bgKK_0{}^0$ the inequality $\mathcal{R}>-1/2$ fails to hold, and, thereby, the hyperbolicity condition relevant for ``modified parabolic-hyperbolic equations'' in \cite{Beyer:2020zlo} is violated, the corresponding solution blows up. In contrast, integrating the parabolic-hyperbolic equations by applying our proposal \eqref{eq:newscR} with setting $\alpha=1.5$, yields---due to the fact that the principal part was retained, and also to the intimate coupling of various modes of the involved variables---a completely regular strongly asymptotically flat initial data such that $\dKK_0{}^0$ decays with the rate $\mathcal{O}(r^{-2})$.

	\medskip

	Note also that there is a significant conceptual difference between using the constraints' parabolic-hyperbolic and algebraic-hyperbolic forms. Whereas the parabolic-hyperbolic system can only be solved in general only on that side of the $\varrho=\varrho_0$ initial data surface where $\Kstar$ is guaranteed to be positive, the algebraic-hyperbolic form of the constraints can always be solved on both sides of the initial data surface, provided that $\kkappa\,\KK<0$. From this point of view, it is notable that \eqref{eq:ahmODE} is deduced from the parabolic form of the Hamiltonian constraint \eqref{eq:phN}. Despite its parabolic origin, \eqref{eq:ahmODE} is an ordinary differential equation for $\Nh_0{}^0$. Thereby, it can be integrated on both sides of the $\varrho=\varrho_0$ initial data surface if it was aimed to solve the algebraic-hyperbolic system on both sides of $\mathscr{S}_{\varrho_0}$. Since, in this paper, the integration starts at $r=1$, and it always happens towards infinity, the above-mentioned favorable aspects of the use of \eqref{eq:ahmODE} remain to economize in future investigations.

	\medskip
	The results covered by the present paper provide significant credit to the investigations in \cite{Racz:2016jej}. It was argued there that in advance of solving the constraints, one could determine each of the ADM charges, the mass, center of mass, and the linear and angular momentum of initial datasets relevant for binary black hole systems. This could be done as the values of the ADM charges are not affected by the particular functional form of the constrained variables. However, we made two critical assumptions in  \cite{Racz:2016jej}. First, even though the explicit form of the constrained variables was irrelevant, their falloff behavior---making them compatible with the strong asymptotic flatness of the yielded complete data $(h_{ab}, K_{ab})$---was assumed in deriving the ADM charges. Second, it was also presumed that the original form of the parabolic-hyperbolic form of the constraints holds. Given these points, it is transparent that the results covered by the present paper provide a significant verification of the assumptions made in \cite{Racz:2016jej}, thus, also to the determination of the ADM charges of initial datasets of binary black hole systems.

  At this point, we remind the reader that although the main concern of the present paper is the near-Kerr initial data of a single black hole, there are parallel studies that verify that the evolutionary formulation of the constraints is capable of producing binary black hole initial data \cite{Beyer:2019kty,Beyer:2020zlo,Nakonieczna:2017eev,Doulis:2019egi}. We believe that the methods presented in this paper are very useful for generating initial data for binary black holes with appropriate asymptotics.  However, the verification of the latter claim is left to future studies.

	\medskip

	Finally, it would be good to know that the novel method introduced in this paper will find its way to be adopted in analytic investigations. It may provide stimulating munition to prove the global existence of solutions to the evolutionary forms of the constraints or derive exact decay rates for various asymptotically flat near-Kerr initial datasets. Applying our new proposal to other more challenging asymptotically flat configurations would also deserve further investigation.

  \section{Acknowledgments}

  The authors would like to thank Florian Beyer and Joshua Ritchie for helpful comments. This work was supported in part by NKFIH Grant No. K-115434. K. C.\,also acknowledges support by NSF CAREER Award No. PHY-2047382.

  \printbibliography

@online{code,
  author = {Csuk\'as, K\'aroly},
  title = {GridRipper webpage},
  url = {http://www.rmki.kfki.hu/~gridripper/},
  urldate = {2023-03-07}
}

@article{Lichnerowicz:1944,
 author = "{Lichnerowicz, Andre}",
 title = "{L'int{\'e}gration des {\'e}quations de la gravitation relativiste et le probl{\`e}me des {{\(n\)}} corps}",
 %fjournal = "{Journal de Math{\'e}matiques Pures et Appliqu{\'e}es. Neuvi{\`e}me S{\'e}rie}",
 journal = "{J. Math. Pures Appl. (9)}",
 %ISSN = "{0021-7824}",
 volume = "{23}",
 pages = "{37--63}",
 year = "{1944}"
}

@article{Lichnerowicz:1952,
     author = "{Lichnerowicz, Andr\'e}",
     title = "{Sur les \'equations relativistes de la gravitation}",
     journal = "{Bulletin de la Soci\'et\'e Math\'ematique de France}",
     pages = "{237--251}",
     publisher =" {Soci\'et\'e math\'ematique de France}",
     volume = "{80}",
     year = "{1952}",
     doi = "{10.24033/bsmf.1433}"
}

@article{York:1972prl,
  title = "{Role of Conformal Three-Geometry in the Dynamics of Gravitation}",
  author = "{York, James W.}",
  journal = "{Phys. Rev. Lett.}",
  volume = "{28}",
  issue = "{16}",
  pages = "{1082--1085}",
  numpages = "{0}",
  year = "{1972}",
  month = "{Apr}",
  publisher =" {American Physical Society}",
  doi = "{10.1103/PhysRevLett.28.1082}",
}

@article{York:1975aihp,
 author =" {York, James W. jun.}",
 title = "{Covariant decompositions of symmetric tensors in the theory of gravitation}",
 %FJournal = "{Annales de l'Institut Henri Poincar{\'e}. Nouvelle S{\'e}rie. Section A. Physique Th{\'e}orique}",
 journal = "{Ann. Inst. Henri Poincar{\'e}, Nouv. S{\'e}r., Sect. A}",
 %ISSN = "{0020-2339}",
 volume = "{21}",
 pages = "{319--332}",
 year = "{1975}"
}

@article{Stachel:1978,
author = {d’Inverno,R. A.  and Stachel,John },
title = {Conformal two‐structure as the gravitational degrees of freedom in general relativity},
journal = {Journal of Mathematical Physics},
volume = {19},
number = {12},
pages = {2447-2460},
year = {1978},
doi = {10.1063/1.523650},
URL = {https://doi.org/10.1063/1.523650}  
}

@article{Bishop:2004,
  title = {Black hole initial data from a nonconformal decomposition},
  author = {Bishop, Nigel T. and Beyer, Florian and Koppitz, Michael},
  journal = {Phys. Rev. D},
  volume = {69},
  issue = {6},
  pages = {064010},
  numpages = {5},
  year = {2004},
  month = {Mar},
  publisher = {American Physical Society},
  doi = {10.1103/PhysRevD.69.064010},
  url = {https://link.aps.org/doi/10.1103/PhysRevD.69.064010}
}

@article{Beyer:2021kmi,
    author = "Beyer, Florian and Ritchie, Joshua",
    title = "{Asymptotically hyperboloidal initial data sets from a parabolic\textendash{}hyperbolic formulation of the Einstein vacuum constraints}",
    eprint = "2104.10290",
    archivePrefix = "arXiv",
    primaryClass = "gr-qc",
    doi = "10.1088/1361-6382/ac79f1",
    journal = "Class. Quant. Grav.",
    volume = "39",
    number = "14",
    pages = "145012",
    year = "2022"
}

@article{Nakonieczna:2017eev,
    author = "Nakonieczna, Anna and Nakonieczny, \L{}ukasz and R\'acz, Istv\'an",
    title = "{Black hole initial data by numerical integration of the parabolic\textendash{}hyperbolic form of the constraints}",
    eprint = "1712.00607",
    archivePrefix = "arXiv",
    primaryClass = "gr-qc",
    doi = "10.1142/S021827182150111X",
    journal = "Int. J. Mod. Phys. D",
    volume = "30",
    number = "15",
    pages = "2150111",
    year = "2021"
}

@article{Doulis:2019egi,
    author = "Doulis, Georgios",
    title = "{Construction of high precision numerical single and binary black hole initial data}",
    eprint = "1906.01479",
    archivePrefix = "arXiv",
    primaryClass = "gr-qc",
    doi = "10.1103/PhysRevD.100.024064",
    journal = "Phys. Rev. D",
    volume = "100",
    number = "2",
    pages = "024064",
    year = "2019"
}

@article{Beyer:2020zlo,
    author = {Beyer, Florian and Frauendiener, J\"org and Ritchie, Joshua},
    title = "{Asymptotically flat vacuum initial data sets from a modified parabolic-hyperbolic formulation of the Einstein vacuum constraint equations}",
    eprint = "2002.06759",
    archivePrefix = "arXiv",
    primaryClass = "gr-qc",
    doi = "10.1103/PhysRevD.101.084013",
    journal = "Phys. Rev. D",
    volume = "101",
    number = "8",
    pages = "084013",
    year = "2020"
}

@article{Chen:2021rtb,
    author = "Chen, Yitian and Deppe, Nils and Kidder, Lawrence E. and Teukolsky, Saul A.",
    title = "{Efficient simulations of high-spin black holes with a new gauge}",
    eprint = "2108.02331",
    archivePrefix = "arXiv",
    primaryClass = "gr-qc",
    doi = "10.1103/PhysRevD.104.084046",
    journal = "Phys. Rev. D",
    volume = "104",
    number = "8",
    pages = "084046",
    year = "2021"
}

@article{Beyer:2019kty,
    author = {Beyer, Florian and Escobar, Leon and Frauendiener, J\"org and Ritchie, Joshua},
    title = "{Numerical construction of initial data sets of binary black hole type using a parabolic-hyperbolic formulation of the vacuum constraint equations}",
    eprint = "1903.06329",
    archivePrefix = "arXiv",
    primaryClass = "gr-qc",
    doi = "10.1088/1361-6382/ab3482",
    journal = "Class. Quant. Grav.",
    volume = "36",
    pages = "175005",
    year = "2019"
}

@article{Racz:2016jej,
    author = "R\'acz, Istv\'an",
    title = "{Can We Prescribe the Physical Parameters of Multiple Black Holes?}",
    eprint = "1608.02283",
    archivePrefix = "arXiv",
    primaryClass = "gr-qc",
    doi = "10.3390/math9243170",
    journal = "Mathematics",
    volume = "9",
    number = "24",
    pages = "3170",
    year = "2021"
}

@article{Beyer:2017njj,
    author = {Beyer, Florian and Escobar, Leon and Frauendiener, J\"org},
    title = "{Asymptotics of solutions of a hyperbolic formulation of the constraint equations}",
    eprint = "1706.06700",
    archivePrefix = "arXiv",
    primaryClass = "gr-qc",
    doi = "10.1088/1361-6382/aa8be6",
    journal = "Class. Quant. Grav.",
    volume = "34",
    number = "20",
    pages = "205014",
    year = "2017"
}

@article{Csukas:2019qco,
    author = "Csuk\'as, K\'aroly and R\'acz, Istv\'an",
    title = "{Numerical investigations of the asymptotics of solutions to the evolutionary form of the constraints}",
    eprint = "1911.02900",
    archivePrefix = "arXiv",
    primaryClass = "gr-qc",
    doi = "10.1088/1361-6382/ab8fce",
    journal = "Class. Quant. Grav.",
    volume = "37",
    number = "15",
    pages = "155006",
    year = "2020"
}

@article{Racz:2017krc,
    author = "R\'acz, Istv\'an and Winicour, Jeffrey",
    title = "{Toward computing gravitational initial data without elliptic solvers}",
    eprint = "1712.03294",
    archivePrefix = "arXiv",
    primaryClass = "gr-qc",
    doi = "10.1088/1361-6382/aac5c5",
    journal = "Class. Quant. Grav.",
    volume = "35",
    number = "13",
    pages = "135002",
    year = "2018"
}

@article{Racz:2015ena,
    author = "R\'acz, Istv\'an and Winicour, Jeffrey",
    title = "{Black hole initial data without elliptic equations}",
    eprint = "1502.06884",
    archivePrefix = "arXiv",
    primaryClass = "gr-qc",
    doi = "10.1103/PhysRevD.91.124013",
    journal = "Phys. Rev. D",
    volume = "91",
    number = "12",
    pages = "124013",
    year = "2015"
}

@article{Racz:2016wcs,
    author = "R\'acz, Istv\'an and Winicour, Jeffrey",
    title = "{On solving the constraints by integrating a strongly hyperbolic system}",
    eprint = "1601.05386",
    archivePrefix = "arXiv",
    primaryClass = "gr-qc",
    month = "1",
    year = "2016"
}

@article{Racz:2015mfa,
    author = "R\'acz, Istv\'an",
    title = "{Constraints as evolutionary systems}",
    eprint = "1508.01810",
    archivePrefix = "arXiv",
    primaryClass = "gr-qc",
    doi = "10.1088/0264-9381/33/1/015014",
    journal = "Class. Quant. Grav.",
    volume = "33",
    number = "1",
    pages = "015014",
    year = "2016"
}

@article{Racz:2014jra,
    author = "R\'acz, Istv\'an",
    title = "{Dynamical determination of the gravitational degrees of freedom}",
    eprint = "1412.0667",
    archivePrefix = "arXiv",
    primaryClass = "gr-qc",
    month = "12",
    year = "2014"
}

@article{Racz:2014gea,
    author = "R\'acz, Istv\'an",
    title = "{Cauchy problem as a two-surface based `geometrodynamics'}",
    eprint = "1409.4914",
    archivePrefix = "arXiv",
    primaryClass = "gr-qc",
    doi = "10.1088/0264-9381/32/1/015006",
    journal = "Class. Quant. Grav.",
    volume = "32",
    pages = "015006",
    year = "2015"
}

@article{Dain:2001ry,
    author = "Dain, Sergio and Friedrich, Helmut",
    title = "{Asymptotically flat initial data with prescribed regularity at infinity}",
    eprint = "gr-qc/0102047",
    archivePrefix = "arXiv",
    doi = "10.1007/s002200100524",
    journal = "Commun. Math. Phys.",
    volume = "222",
    pages = "569--609",
    year = "2001"
}

@article{Chrusciel:1986xts,
    author = "Chrusciel, Piotr T.",
    editor = "Bergmann, P. G. and De Sabbata, V.",
    title = "{Boundary Conditions at Spatial Infinity}: {From a Hamiltonian Point of View}",
    eprint = "1312.0254",
    archivePrefix = "arXiv",
    primaryClass = "gr-qc",
    doi = "10.1007/978-1-4899-3626-4_5",
    journal = "NATO Sci. Ser. B",
    volume = "138",
    pages = "49--59",
    year = "1986"
}

@article{Johansson:2015cca,
    author = "Johansson, H. T. and Forss\'en, C.",
    title = "{Fast and accurate evaluation of Wigner 3j, 6j, and 9j symbols using prime factorisation and multi-word integer arithmetic}",
    eprint = "1504.08329",
    archivePrefix = "arXiv",
    primaryClass = "physics.comp-ph",
    doi = "10.1137/15M1021908",
    journal = "SIAM J. Sci. Statist. Comput.",
    volume = "38",
    pages = "A376--A384",
    year = "2016"
}

@book{Choquet-Bruhat:2014okh,
    author = "Choquet-Bruhat, Yvonne",
    title = "{Introduction to General Relativity, Black Holes, and Cosmology}",
    isbn = "978-0-19-966645-4, 978-0-19-966646-1",
    publisher = "Oxford University Press",
    month = "11",
    year = "2014"
}

@book{Wald:1984rg,
    author = "Wald, Robert M.",
    title = "{General Relativity}",
    doi = "10.7208/chicago/9780226870373.001.0001",
    publisher = "Chicago Univ. Pr.",
    address = "Chicago, USA",
    year = "1984"
}

\end{document}